\documentclass[preprint, 12pt]{elsarticle}

\usepackage{fullpage}
\usepackage{array}
\usepackage{algorithm}
\usepackage{algpseudocode}
\usepackage{pifont}
\usepackage{bbding}
\usepackage{graphicx}
\usepackage{subfig}
\usepackage{enumerate}
\usepackage[shortlabels]{enumitem}
\usepackage{amsmath,amssymb,mathtools}
\usepackage{amsfonts,bm}
\usepackage{todonotes}
\usepackage[commentmarkup=footnote]{changes}
\usepackage{stmaryrd}
\usepackage{siunitx}
\usepackage{caption}
\usepackage{array}
\usepackage{booktabs}
\usepackage{makecell}
\usepackage{nomencl}
\usepackage{lineno}
\usepackage{url}
\usepackage{natbib}
\biboptions{sort&compress}
\usepackage[strings]{underscore}
\usepackage{tikz}
\usetikzlibrary{shapes.geometric}
\usepackage{xcolor}

\definecolor{RoseMagenta}{RGB}{160, 0, 80}

\DeclareRobustCommand{\redcross}[1][1.0]{%
\tikz[baseline=-0.5ex, scale=#1, line width=1.2pt, RoseMagenta]{
    \draw (-0.25em,-0.25em) -- (0.25em,0.25em);
    \draw (-0.25em,0.25em) -- (0.25em,-0.25em);
}}

\DeclareRobustCommand{\blackstar}[1][1.0]{%
\tikz[baseline=-0.5ex, scale=#1]{
    \node[star, star points=5, star point ratio=2.25,
          fill=black, draw=black, inner sep=0.9pt] at (0,0) {};
}}

\makeatletter
\def\ps@pprintTitle{%
  \let\@oddhead\@empty
  \let\@evenhead\@empty
  \def\@oddfoot{\hfil\thepage\hfil}%
  \let\@evenfoot\@oddfoot
}
\makeatother

\graphicspath{{./figs/}}

\begin{document}

\begin{frontmatter}

\title{Ensemble generative filtering for sequential data assimilation in dynamical systems}

\author[sd,cal]{Xu-Hui Zhou}%
\author[fi]{Jiequn Han\corref{cor1}}%

\ead{jhan@flatironinstitute.org}

\cortext[cor1]{Corresponding author.}

\affiliation[sd]{organization={Scripps Institution of Oceanography, University of California San Diego},
            %addressline={}, 
            city={La Jolla},
            postcode={92093}, 
            state={CA},
            country={USA}}
\affiliation[cal]{organization={Computing + Mathematical Sciences, California Institute of Technology},
            %addressline={}, 
            city={Pasadena},
            postcode={91125}, 
            state={CA},
            country={USA}}
\affiliation[fi]{organization={Center for Computational Mathematics, Flatiron Institute},
            %addressline={}, 
            city={New York},
            postcode={10010}, 
            state={NY},
            country={USA}}

\begin{abstract}
Sequential data assimilation (DA) faces a fundamental trade-off:
particle filters capture non-Gaussian cycling priors but require prohibitively large ensembles, whereas the ensemble Kalman filter (EnKF) is computationally efficient but constrained by its Gaussian assumption.
As machine learning enables rapid model forecasts, exploiting non-Gaussian prior features via moderately large ensembles has become increasingly viable.
To exploit this opportunity, we propose the ensemble generative filter (EnGF), a simple yet effective method for non-Gaussian filtering.
The key idea is to fit a generative model to the forecast ensemble at each DA cycle and harness its defining strength, inexpensive sampling, to draw a much larger particle population for Bayesian analysis without any additional model forecasts; we adopt a Gaussian mixture model as a lightweight instance that can be fit cheaply from a moderate ensemble.
To address practical challenges, we further extend the EnGF by introducing (i) a tempered EnGF using likelihood tempering to prevent particle degeneracy under informative observations and (ii) a latent EnGF that performs prior modeling and Bayesian updates in a reduced latent space for high-dimensional systems.
Across chaotic systems (doubling map, Lorenz-63, and Lorenz-96) and a challenging shock-tube problem, the EnGF delivers clear and often substantial improvements over the EnKF, in some cases even approaching the filtering accuracy of a massive-ensemble particle filter at a small fraction of its cost.
\end{abstract}


\end{frontmatter}

% \linenumbers

%% main text
\section{Introduction}
\label{sec:introduction}

\subsection{Background and motivation}

Data assimilation (DA) provides a systematic framework for updating model predictions using observational information~\cite{law2015data}.
Mathematically, sequential DA can be viewed as recursively updating a probability distribution over the system state according to Bayes' rule as new observations become available.
Numerical DA methods can therefore be interpreted as different computational approximations of this Bayesian updating process.
Among these methods, particle filters (PFs) seek to sample the full filtering posterior without imposing a parametric form~\cite{evensen2022data}, thereby accommodating general, potentially non-Gaussian distributions.
However, PFs suffer from ensemble degeneracy in high-dimensional systems and can require prohibitively large ensembles~\cite{snyder2008obstacles}.
This requirement is especially restrictive for complex physical systems, where the computational cost of forward simulations typically limits practical ensemble sizes to $O(10)$–$O(10^2)$ members~\cite{miyoshi201410,chattopadhyay2023deep,chan2024improving,tsyrulnikov2024regularization}.

Given these practical constraints, the ensemble Kalman filter (EnKF) has become a dominant approach for high-dimensional DA owing to its computational tractability~\cite{evensen1994sequential,evensen2009data}, with widespread applications in numerical weather prediction, physical oceanography, reservoir engineering, and other scientific and engineering fields (see, e.g.,~\cite{houtekamer2016review,evensen1996assimilation,gu2005history,zhou2023inference,kaveh2026data,zafar2026data,jain2026parasitic}).
The EnKF makes use of an implicit Gaussian approximation of the filtering prior, enabling a closed-form analysis update based only on the mean and covariance estimated from the forecast ensemble.
This approximation is particularly reasonable when the ensemble size is small, as higher-order features of the filtering prior generally cannot be reliably characterized from a small number of samples.
However, recent advances in reduced-order modeling and machine learning, together with increasing computational power, are making larger forecast ensembles feasible~\cite{bi2023accurate,kurth2023fourcastnet,lam2023learning,lapillonne2026operational}, potentially expanding the computationally affordable regime from historically small ensembles to moderately large ones (e.g., $O(10^2)$--$O(10^3)$ members, or beyond)~\cite{alet2026operational,zeba2026anomalous}.
While larger ensembles can effectively reduce sampling error, their benefit to the EnKF may eventually plateau because the Gaussian approximation remains a limiting factor for non-Gaussian filtering priors~\cite{spantini2022coupling}.
This raises a natural question: \textit{Can moderately large ensembles be leveraged to move beyond the Gaussian approximation, thereby more faithfully representing the filtering prior and improving DA performance?}

\subsection{Literature review and context}

The idea of relaxing the Gaussian assumption to improve EnKF performance is not new.
A natural strategy is to combine the EnKF with PF to retain EnKF robustness while better accommodating non-Gaussian filtering distributions.
One example is the weighted ensemble Kalman filter, which uses an EnKF-based update to construct an observation-informed proposal distribution and subsequently applies importance weighting so that the weighted particles represent the target posterior~\cite{papadakis2010data}.
Another representative approach is the ensemble Kalman particle filter, which splits the likelihood into two parts, with the first assimilated using an EnKF update and the second incorporated through particle weighting and resampling~\cite{frei2013bridging}.
Although the Kalman-type update helps mitigate particle degeneracy by guiding particles toward regions favored by the observations, the Gaussian approximation underlying this update can weaken non-Gaussian features of the filtering prior before the subsequent particle-based correction.

Another line of research seeks to represent the filtering prior using more flexible density models.
Kernel density estimation is one such strategy, constructing a continuous approximation of the filtering prior by associating each particle with a local kernel.
When Gaussian kernels are used, the resulting density takes the form of a Gaussian mixture model (GMM) with one component centered at each particle.
Based on this representation, the low-rank kernel particle Kalman filter assimilates observations by applying a Kalman-type update to each Gaussian component, adjusting its mixture weight, and resampling the resulting mixture to form the analysis ensemble~\cite{hoteit2008new}.
This strategy was further extended through adaptive adjustment of the mixture weighting to mitigate weight degeneracy~\cite{stordal2011bridging}.
However, the resulting prior representation can be sensitive to the choice of kernel covariance, which must be appropriately specified for the forecast ensemble.

Alternatively, GMMs can be fitted directly to the forecast ensemble, with the mixture components inferred from the ensemble rather than centered at individual particles. 
These approaches~\cite{smith2007cluster,sun2009sequential,sondergaard2013dataI,sondergaard2013dataII} typically use the expectation–maximization (EM) algorithm to fit the GMM, with the number of components selected using criteria such as the Akaike information criterion (AIC) or Bayesian information criterion (BIC). 
During the analysis, each Gaussian component is updated separately, and the resulting mixture is used to construct the analysis ensemble. 
However, dividing a finite forecast ensemble into multiple components leaves fewer samples for estimating the covariance of each component, which can lead to poorly conditioned covariance estimates, particularly in high-dimensional systems. 
Moreover, for nonlinear observation models, the component-wise Gaussian update is generally no longer available in closed form and requires additional approximations.

More recently, generative modeling in machine learning has provided more flexible tools for representing the filtering prior and generating particles guided by observations~\cite{hodyss2026using,morzfeld2026condition}.
A representative example is the score-based filter, which characterizes the filtering prior through its score function and generates analysis particles through reverse-time diffusion~\cite{bao2024score}.
However, the score model requires repeated training with large forecast ensembles across DA cycles, leading to substantial computational cost.
To eliminate this training requirement, the ensemble score filter (EnSF) directly approximates the score function from the finite forecast ensemble using a Monte Carlo estimator~\cite{bao2024ensemble}.
Subsequent extensions of the EnSF introduced an iterative formulation to improve posterior score estimation~\cite{zhang2025iensf} and a two-step formulation for partially observed systems~\cite{xiong2026two}.
Other recent generative approaches to sequential DA include, for example, closed-form conditional diffusion~\cite{binder2026closed}, flow matching~\cite{transue2025flow}, and hybrid ensemble--score methods~\cite{behnoudfar2026omega}.
Across these generative filters, the analysis relies on learned or Monte Carlo approximations whose errors are difficult to quantify and control, and the resulting filters generally lack theoretical guarantees~\cite{bruna2024posterior,zhang2025iensf,ren2025driftlite}.

Another related but distinct direction is learning-based ensemble filtering, which uses neural operators, often motivated by a mean-field formulation, to learn an analysis map directly from the forecast ensemble and observations to the analysis ensemble~\cite{zhou2024bi,bach2026learning}.
Once trained, the map can be reused across DA cycles without assuming an explicit parametric form for the filtering distribution, but it is typically tied to its training configuration and may require retraining when the DA setup, such as the observation operator, changes.
Current implementations of these methods also rely on supervised learning, requiring either high-quality reference analysis ensembles or synthetic trajectories with known truth, 
which are rarely available or fully faithful in real applications.
Overall, these approaches follow a different methodological paradigm from the present work.

\subsection{Proposed approach and paper outline}

In this work, we introduce the ensemble generative filter (EnGF).
The key idea is to exploit the defining capability of a generative model, cheap sampling, to enrich the Bayesian analysis: at each DA cycle, we fit a generative model to the moderately large forecast ensemble and draw from it a much larger particle population, raising the Monte Carlo resolution of the analysis without any additional model forecasts.
The analysis is performed through likelihood weighting and resampling to the original ensemble size for the next DA cycle.

Compared with PFs, the EnGF enables Bayesian analysis with a much larger particle population without propagating the additional particles through the dynamical model, while, compared with the EnKF, it can exploit richer features of the filtering prior beyond those captured by a single Gaussian approximation.
In contrast to existing mixture-based filters, which assimilate observations through Kalman-type updates on individual mixture components, and to more expressive score-based diffusion or flow-matching filters, which rely on a learned score or transport, the EnGF analysis requires no covariance inversion or posterior approximation, and it applies directly to nonlinear observation operators.
A further appeal of the EnGF is its simplicity, both conceptual and practical: the core method pairs a lightweight generative-model fit with importance weighting and resampling, requires no model derivatives or expensive learning components, and is straightforward to implement on top of an existing ensemble forecasting system.

More abstractly, the EnGF only requires a generative model that is (i) cheap to fit from a moderate ensemble of $O(10^2)$--$O(10^3)$ samples, (ii) cheap to draw many samples from, and (iii) expressive enough to improve on a single Gaussian.
We adopt the GMM as a lightweight instance satisfying all three: it can be fitted from finite ensembles by well-established EM procedures, sampled at negligible cost, and adapted at each DA cycle, a combination that most more expressive generative models sacrifice.
Other generative models could be substituted as this trade-off evolves.
In practice, however, the basic EnGF faces two challenges: particle degeneracy under informative observations and effective prior modeling in high-dimensional systems, which remains fundamentally difficult even with moderately large ensembles.
We mitigate these challenges through two simple extensions, the tempered EnGF and the latent EnGF, while keeping the core generative filtering algorithm unchanged.

These design choices translate into strong empirical performance.
Notably, in Lorenz-63, the EnGF closes nearly $80\%$ of the filtering-error gap between the EnKF and a massive-ensemble PF using less than $0.1\%$ of the particles required by the PF.
In the shock-tube problem, a challenging non-Gaussian regime with discontinuous states and sparse observations, the latent EnGF accurately recovers sharp flow features and produces analysis states that closely and consistently track the true dynamics.

The remainder of this paper is organized as follows.
Section~\ref{sec:preliminaries} introduces the preliminaries of ensemble-based filtering and briefly reviews the PF and the EnKF.
Section~\ref{sec:method} presents the EnGF and its tempered and latent extensions.
Section~\ref{sec:result} evaluates the proposed methods through a series of numerical experiments.
Finally, Section~\ref{sec:conclusion} summarizes the main conclusions.

\section{Preliminaries}
\label{sec:preliminaries}

\subsection{Ensemble-based filtering}

\begin{figure}[!htb]
\centering
\includegraphics[width=0.86\textwidth]{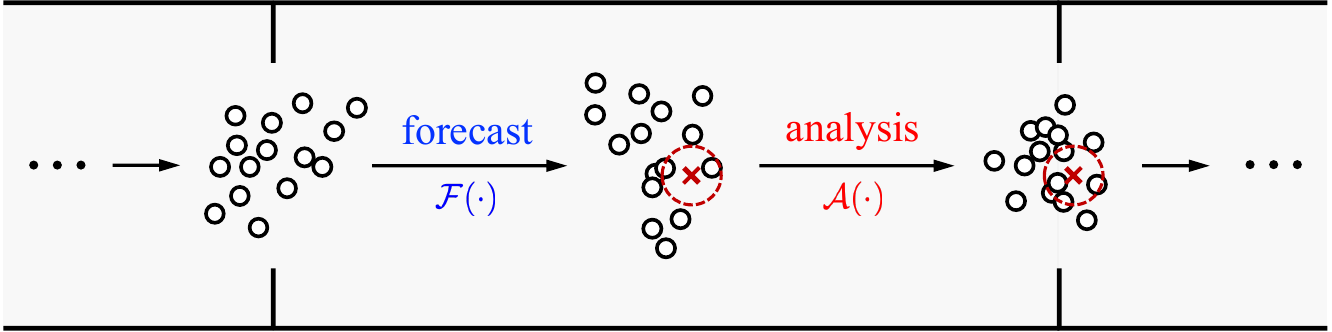}
\caption{Schematic of a generic DA cycle consisting of a \textit{forecast} step, in which each particle is propagated independently through the forecast operator $\mathcal F$, and an \textit{analysis} step, in which the resulting forecast ensemble is mapped to the analysis ensemble through the analysis operator $\mathcal A$, conditioned on the observation (denoted by \redcross{}).}
\label{fig:ensemble-filtering}
\end{figure}

We focus here on ensemble-based filtering methods for sequential DA, in which the state distribution is approximated by a finite ensemble of particles.
As observations are assimilated sequentially, the ensemble is updated through a sequence of DA cycles.
Fig.~\ref{fig:ensemble-filtering} illustrates a generic DA cycle, consisting of two steps: forecast and analysis.
Since the same procedure is repeated in each DA cycle, we omit temporal indices throughout this paper.

In the forecast step, each particle is propagated independently through a forecast operator, denoted by $\mathcal{F}$, typically by numerically integrating the governing equations.
The resulting ensemble provides a particle approximation of the filtering prior distribution $\pi^{\mathrm f}(\mathbf{z})$ and is written as
\begin{equation}
    \mathbf{Z}^{\mathrm f}
    =
    \left[
    \mathbf{z}^{\mathrm f}_1,\,
    \mathbf{z}^{\mathrm f}_2,\,
    \ldots,\,
    \mathbf{z}^{\mathrm f}_{n_e}
    \right]
    \in \mathbb{R}^{n_z \times n_e},
    \label{eq:forecast-ensemble}
\end{equation}
where $\mathbf{z}^{\mathrm f}_i$ denotes the $i$th forecast particle, and $n_z$ and $n_e$ denote the state dimension and ensemble size, respectively.

In the analysis step, an observation $\mathbf{d} \in \mathbb{R}^{n_d}$ is incorporated through the observation model
\begin{equation}
    \mathbf{d}
    =
    \mathcal{H}(\mathbf{z})
    +
    \boldsymbol{\eta},
    \qquad
    \boldsymbol{\eta}
    \sim
    \mathcal{N}(\mathbf{0},\mathbf{R}),
    \label{eq:observation-model}
\end{equation}
where $\mathcal{H}: \mathbb{R}^{n_z} \rightarrow \mathbb{R}^{n_d}$ denotes the observation operator that maps the state to the observation space, and $\mathbf{R} \in \mathbb{R}^{n_d \times n_d}$ denotes the observation-error covariance matrix (a symmetric positive semidefinite matrix).
The filtering prior is then conditioned on the observation $\mathbf{d}$ to obtain the filtering posterior $\pi^{\mathrm a}(\mathbf{z})$, which, by Bayes' rule, satisfies
\begin{equation}
    \pi^{\mathrm a}(\mathbf{z})
    \propto
    p(\mathbf{d}\mid\mathbf{z})\,
    \pi^{\mathrm f}(\mathbf{z}),
    \label{eq:bayesian-analysis}
\end{equation}
where $p(\mathbf{d}\mid\mathbf{z})$ denotes the likelihood.
For a finite ensemble, Eq.~\eqref{eq:bayesian-analysis} is realized approximately through an analysis operator $\mathcal{A}$, which maps the forecast ensemble to an analysis ensemble,
\begin{equation}
    \mathbf{Z}^{\mathrm a}
    =
    \mathcal{A}(\mathbf{Z}^{\mathrm f};\mathbf{d}).
    \label{eq:analysis-operator}
\end{equation}

In the following, we review two representative constructions of the analysis operator $\mathcal{A}$, the particle filter (PF) and the ensemble Kalman filter (EnKF), whose complementary limitations motivate the ensemble generative filter (EnGF) introduced in Sec.~\ref{sec:method}.

\subsection{Particle filter (PF)}
The PF represents the filtering posterior directly as a weighted forecast ensemble. Take the classical bootstrap PF~\cite{gordon1993novel} as an example, in which resampling at every DA cycle keeps the forecast particles
$\{\mathbf z_i^{\mathrm f}\}_{i=1}^{n_e}$ equally weighted. Each particle is then
assigned a weight proportional to its observation likelihood,
\begin{equation}
    w_i\propto
    \exp\!\left[
        -\frac{1}{2}
        \left(\mathbf d-\mathcal H(\mathbf z_i^{\mathrm f})\right)^\top
        \mathbf R^{-1}
        \left(\mathbf d-\mathcal H(\mathbf z_i^{\mathrm f})\right)
    \right],
    \qquad i=1,\ldots,n_e,
    \label{eq:pf-weights}
\end{equation}
giving $\pi^{\mathrm a}(\mathbf z)\approx\sum_{i} w_i\,
\delta(\mathbf z-\mathbf z_i^{\mathrm f})$ after normalization.\footnote{In general
$w_i\propto w_i^{\mathrm f}\,p(\mathbf d\mid\mathbf z_i^{\mathrm f})$, reducing to Eq.~\eqref{eq:pf-weights} for uniform prior weights $w_i^{\mathrm f}$.} To avoid weight degeneracy, the ensemble is resampled into an equally weighted analysis ensemble \cite{doucet2009tutorial}.

As the ensemble size grows, the PF converges to the true filtering distribution, making it a high-fidelity reference (when computationally affordable) for assessing other methods. 
However, its cost is high: avoiding weight degeneracy typically requires a large ensemble, and every particle must be propagated through the dynamics $\mathcal F$, so the number of forward simulations per DA cycle can become prohibitive.
Reducing this forward-simulation cost is one central motivation for the EnGF developed below.

\subsection{Ensemble Kalman filter (EnKF)}
\label{sec:EnKF}

The EnKF makes use of a Gaussian approximation of the filtering prior, with the required statistics estimated directly from the forecast ensemble $\{\mathbf z_i^{\mathrm f}\}_{i=1}^{n_e}$.
Specifically, the EnKF operates as follows.
The forecast mean and normalized ensemble-anomaly matrix are given by
\begin{equation}
    \overline{\mathbf z}^{\mathrm f}
    =
    \frac{1}{n_e}\sum_{i=1}^{n_e}\mathbf z_i^{\mathrm f},
    \qquad
    \mathbf A_z
    =
    \frac{1}{\sqrt{n_e-1}}
    \left[
        \mathbf z_1^{\mathrm f}-\overline{\mathbf z}^{\mathrm f},
        \ldots,
        \mathbf z_{n_e}^{\mathrm f}-\overline{\mathbf z}^{\mathrm f}
    \right]
    \in \mathbb{R}^{n_z \times n_e}.
\end{equation}

Applying the observation operator to each forecast particle gives the predicted observations
\begin{equation}
    \mathbf y_i
    =
    \mathcal H(\mathbf z_i^{\mathrm f}),
    \qquad i=1,\ldots,n_e.
\end{equation}
The corresponding mean and normalized ensemble-anomaly matrix are
\begin{equation}
    \overline{\mathbf y}
    =
    \frac{1}{n_e}\sum_{i=1}^{n_e}\mathbf y_i,
    \qquad
    \mathbf A_y
    =
    \frac{1}{\sqrt{n_e-1}}
    \left[
        \mathbf y_1-\overline{\mathbf y},
        \ldots,
        \mathbf y_{n_e}-\overline{\mathbf y}
    \right]
    \in \mathbb{R}^{n_d \times n_e}.
\end{equation}

Using these ensemble statistics, the stochastic EnKF updates each forecast particle according to
\begin{equation}
    \mathbf z_i^{\mathrm a}
    =
    \mathbf z_i^{\mathrm f}
    +
    \mathbf A_z\mathbf A_y^\top
    \left(
        \mathbf A_y\mathbf A_y^\top+\mathbf R
    \right)^{-1}
    \left(
        \mathbf d+\boldsymbol{\eta}_i-\mathbf y_i
    \right),
    \qquad
    \boldsymbol{\eta}_i
    \sim
    \mathcal N(\mathbf 0,\mathbf R).
    \label{eq:enkf-update}
\end{equation}
The updated particles $\{\mathbf z_i^{\mathrm a}\}_{i=1}^{n_e}$ form the analysis ensemble.

\begin{figure}[!htb]
\centering
\includegraphics[width=0.99\textwidth]{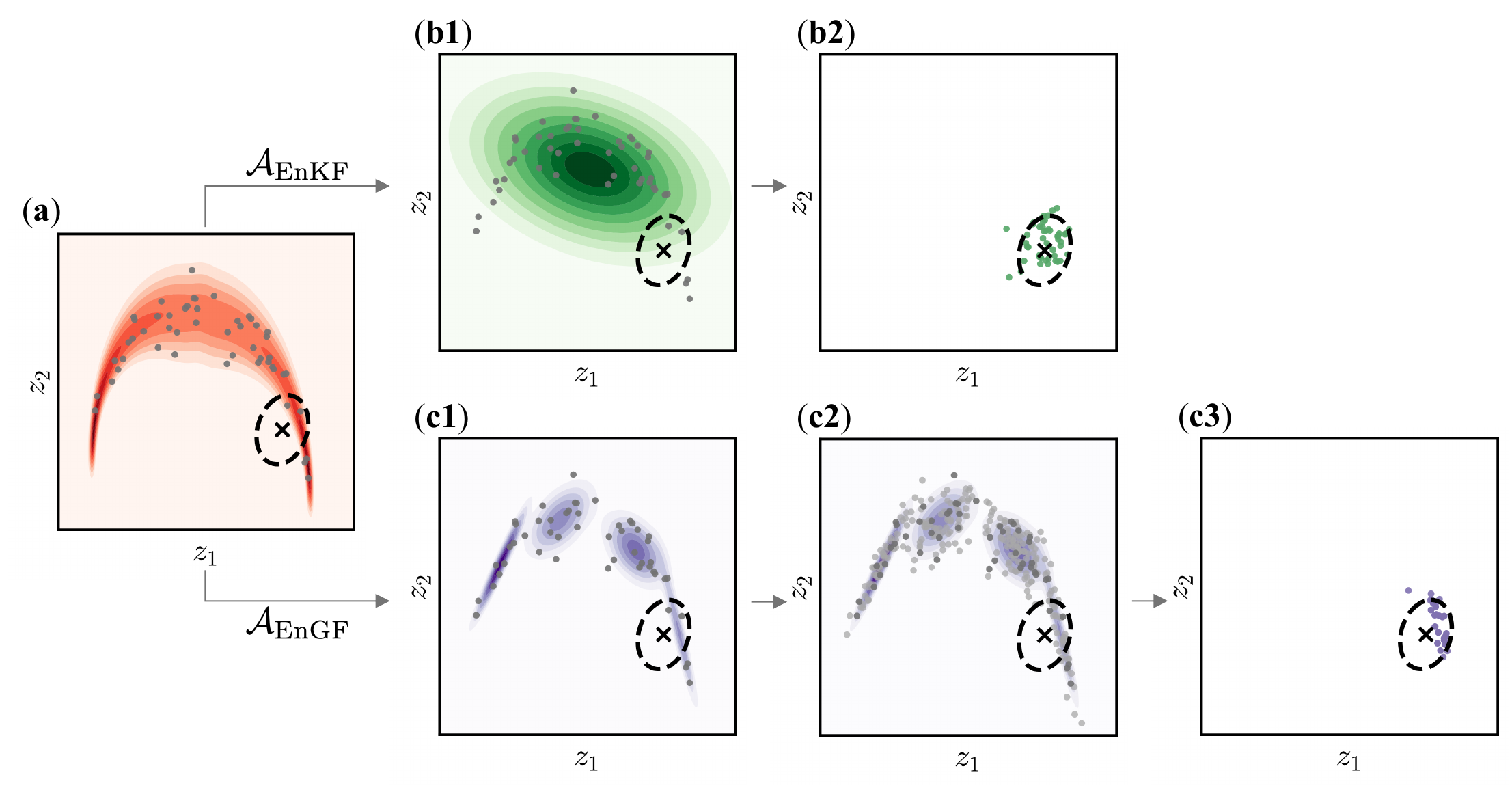}
  \caption{
    Schematic illustration of the EnKF and EnGF analysis steps.
    In (a), gray dots denote the forecast ensemble
    $\{\mathbf z_i^{\mathrm f}\}_{i=1}^{n_e}$,
    and the background density illustrates the underlying filtering prior
    $\pi^{\mathrm f}(\mathbf z)$, which is typically unknown and may be non-Gaussian.
    For the EnKF, (b1) schematically illustrates the implicit Gaussian assumption on the filtering prior rather than an actual computational step, while
    (b2) applies the EnKF update (Eq.~\eqref{eq:enkf-update}) to obtain the analysis ensemble.
    The EnGF (c1) represents the filtering prior using a fitted GMM,
    (c2) generates additional particles (light gray dots) from the GMM to augment the forecast ensemble, and
    (c3) applies systematic resampling according to the likelihood-based weights to obtain the analysis ensemble.
    Both methods retain an analysis ensemble size of $n_e$.
    }
  \label{fig:enkf_engf_schematic}
\end{figure}

The EnKF relies on an implicit Gaussian approximation (Fig.~\ref{fig:enkf_engf_schematic}(b1)) of the filtering prior, which itself may be non-Gaussian (Fig.~\ref{fig:enkf_engf_schematic}(a)).
Under a linear observation operator, the EnKF can remain effective when the forecast dynamics $\mathcal{F}$ are (i) mildly nonlinear, so that the filtering prior is nearly Gaussian, or (ii) moderately nonlinear, so that the filtering prior may be non-Gaussian but the regularizing effect of the observations can render the filtering posterior nearly Gaussian~\cite{morzfeld2019gaussian}.
For strongly nonlinear problems, however, both the filtering prior and posterior can become severely non-Gaussian, and the EnKF performance can deteriorate.
This limitation can be further exacerbated by some nonlinear observation operators.

\section{Methodology}
\label{sec:method}

\subsection{Ensemble generative filter (EnGF)}
\label{sec:EnGF}
The PF and the EnKF represent two extremes.
The PF makes no Gaussian approximation and is asymptotically exact, but attaining a reliable approximation requires a large ensemble, and hence a large number of forward simulations of the dynamics $\mathcal F$, which is typically the dominant computational cost.
The EnKF, in contrast, attains low cost with a small forecast ensemble but relies on a Gaussian approximation that degrades for strongly non-Gaussian problems.

The key idea here is to fit a Gaussian mixture model to the moderate forecast ensemble and then \emph{generate} additional particles from this fitted model without requiring additional model propagation. The expensive dynamical propagation is thus performed only for the $n_e$ forecast particles, while the resampling that realizes the Bayesian update acts on a much larger, cheaply
generated particle set. 
In this way, the EnGF obtains the statistical richness of a large particle ensemble at the dynamical cost of a small one.

As illustrated in Fig.~\ref{fig:enkf_engf_schematic}(c1)--(c3), the generative filtering procedure of the EnGF consists of three main steps:
(i) constructing a GMM representation of the filtering prior,
(ii) generating additional particles from the GMM, and
(iii) incorporating the observation to resample the particles and form the analysis ensemble, which retains the same ensemble size as the forecast ensemble.
These steps are detailed as follows.

\textbf{Step (i): GMM representation.}
The EnGF represents the filtering prior $\pi^{\mathrm f}$ using a GMM,
\begin{equation}
    \pi^{\mathrm f}(\mathbf z)
    =
    \sum_{\ell=1}^{K}
    \phi_\ell
    \mathcal N
    \left(
        \mathbf z;
        \boldsymbol{\mu}_\ell,
        \mathbf\Sigma_\ell
    \right),
    \qquad
    \phi_\ell \geq 0,
    \qquad
    \sum_{\ell=1}^{K}\phi_\ell=1,
    \label{eq:engf-gmm}
\end{equation}
where $K$ is the number of Gaussian components,
and $\phi_\ell$, $\boldsymbol{\mu}_\ell$, and $\mathbf\Sigma_\ell$ denote the weight, mean, and covariance of the $\ell$th component, respectively.
The parameters $\{\widehat{\phi}_\ell, \widehat{\boldsymbol{\mu}}_\ell, \widehat{\mathbf\Sigma}_\ell\}_{\ell=1}^{K}$ are estimated from the forecast ensemble $\{\mathbf z_i^{\mathrm f}\}_{i=1}^{n_e}$ using the expectation--maximization (EM) algorithm, and the number of components $K$ is chosen to minimize the Bayesian information criterion (BIC),
\begin{equation}
\begin{aligned}
    \mathrm{BIC}(K) &= -2\sum_{i=1}^{n_e}\log\left[\sum_{\ell=1}^{K}\widehat{\phi}_{\ell}\,\mathcal N\!\left(\mathbf z_i^{\mathrm f}; \widehat{\boldsymbol{\mu}}_{\ell}, \widehat{\mathbf\Sigma}_{\ell}\right)\right] + n_{\theta,K}\log n_e,\\
    K &= \underset{1\leq K' \leq K_{\max}}{\arg\min}\ \mathrm{BIC}(K'),
\end{aligned}
    \label{eq:engf-bic}
\end{equation}
where $n_{\theta,K}=(K-1)+Kn_z+Kn_z(n_z+1)/2$ is the number of free parameters in the $K$-component GMM when full covariances are used.

This procedure enables adaptive representation of the filtering prior at each DA cycle, allowing the GMM representation to adjust to the evolving forecast ensemble rather than imposing a Gaussian representation in all cases.

\textbf{Step (ii): Particle generation.}
Based on the fitted GMM, $n_g$ additional particles are independently generated according to
\begin{equation}
    \widehat{\mathbf z}_j
    \sim
    \sum_{\ell=1}^{K}
    \widehat{\phi}_\ell
    \mathcal N
    \left(
        \mathbf z;
        \widehat{\boldsymbol{\mu}}_\ell,
        \widehat{\mathbf\Sigma}_\ell
    \right),
    \qquad
    j=1,\ldots,n_g.
    \label{eq:engf-generated-particles}
\end{equation}
The generated particles are combined with the original forecast ensemble to form an augmented particle set,
\begin{equation}
    \left\{
        \widetilde{\mathbf z}_j
    \right\}_{j=1}^{n_p}
    =
    \left\{
        \mathbf z_i^{\mathrm f}
    \right\}_{i=1}^{n_e}
    \cup
    \left\{
        \widehat{\mathbf z}_j
    \right\}_{j=1}^{n_g},
    \qquad
    n_p=n_e+n_g.
    \label{eq:engf-augmented-particles}
\end{equation}
The augmented particle set enriches the representation of the filtering prior while retaining the original forecast ensemble.
In this work, we set $n_g=10n_e$ throughout the numerical experiments.

\textbf{Step (iii): Resampling.}
For each particle in the augmented particle set, the corresponding weight satisfies
\begin{equation}
    \widetilde w_j
    \propto
    \exp
    \left[
        -\frac{1}{2}
        \left(
            \mathbf d-\mathcal H(\widetilde{\mathbf z}_j)
        \right)^\top
        \mathbf R^{-1}
        \left(
            \mathbf d-\mathcal H(\widetilde{\mathbf z}_j)
        \right)
    \right],
    \qquad
    j=1,\ldots,n_p.
    \label{eq:engf-unnormalized-weights}
\end{equation}
The weights are then normalized as
$w_j=\widetilde w_j/\sum_{m=1}^{n_p}\widetilde w_m$.
The analysis ensemble
$\{\mathbf z_i^{\mathrm a}\}_{i=1}^{n_e}$
is obtained by systematic resampling from the augmented particle set according to the normalized weights
$\{w_j\}_{j=1}^{n_p}$.

Together, steps (i)--(iii) form the core generative filtering update of the EnGF.
To keep this update robust over repeated DA cycles, we introduce an EnKF fallback that replaces the generative update in two situations: when the effective sample size (ESS) of the augmented particle set, $N_{\mathrm{eff}}=1/\sum_{j=1}^{n_p}w_j^2$, falls below $n_e$, indicating too few effective particles for reliable resampling; and, for a linear observation operator, when the fitted GMM reduces to a single Gaussian ($K=1$), for which the EnKF analysis is already optimal.
As such, the EnGF has three possible operating modes for a linear observation operator: EnKF fallback for $K=1$, EnKF fallback due to insufficient ESS, and the generative filtering update, whereas for a nonlinear observation operator, only the latter two modes apply. The complete EnGF analysis step, including the two EnKF fallback modes, is summarized in Algorithm~\ref{alg:engf}.

\begin{algorithm}[!htb]
\caption{EnGF analysis step (one DA cycle)}
\label{alg:engf}
\begin{algorithmic}[1]
\Require forecast ensemble $\{\mathbf z_i^{\mathrm f}\}_{i=1}^{n_e}$, observation $\mathbf d$
\Ensure analysis ensemble $\{\mathbf z_i^{\mathrm a}\}_{i=1}^{n_e}$
\State Fit a GMM to the forecast ensemble, with $K$ selected by BIC. \Comment{Step (i)}
\State \textbf{if} $K=1$ and $\mathcal H$ is linear \textbf{then} return the EnKF update.  \Comment{Gaussian--linear fallback}
\State Draw $n_g$ particles from the GMM to form the augmented set $\{\widetilde{\mathbf z}_j\}_{j=1}^{n_p}$.   \Comment{Step (ii)}
\State Weight the augmented set by the likelihood and compute $N_{\mathrm{eff}}$. \Comment{Step (iii)}
\State \textbf{if} $N_{\mathrm{eff}}<n_e$ \textbf{then} return the EnKF update. \Comment{insufficient-ESS fallback}
\State Systematically resample $n_e$ particles to obtain $\{\mathbf z_i^{\mathrm a}\}_{i=1}^{n_e}$.
\end{algorithmic}
\end{algorithm}

The basic EnGF in Algorithm~\ref{alg:engf} is most effective when generative filtering remains its dominant mode. Two scenarios, however, can undermine this:
\begin{itemize}
    \item \textit{Informative observations.}
    A large observation dimension and/or small observation errors can cause severe concentration of the particle weights~\cite{lee2016state}, leading to insufficient ESS and frequent EnKF fallback.

    \item \textit{High-dimensional state spaces.}
    As the state dimension increases, the BIC may increasingly favor a single Gaussian component ($K=1$)~\cite{bhattacharya2014lasso,bouveyron2014model}, resulting in frequent EnKF fallback for a linear observation operator.
\end{itemize}
To mitigate these limitations, we extend the EnGF to the \emph{tempered EnGF} (Sec.~\ref{sec:tempered-EnGF}) and \emph{latent EnGF} (Sec.~\ref{sec:latent-EnGF}), respectively; both reuse the analysis step of Algorithm~\ref{alg:engf} with only minor modifications.

\subsection{Tempered EnGF}
\label{sec:tempered-EnGF}
The tempered EnGF addresses informative observations by incorporating the observation likelihood gradually over multiple stages.
Specifically, the Bayesian analysis update in Eq.~\eqref{eq:bayesian-analysis} is decomposed into a sequence of tempered updates,
\begin{equation}
    \pi^{(s)}(\mathbf z)
    \propto
    p(\mathbf d\mid\mathbf z)^{\alpha_s}
    \pi^{(s-1)}(\mathbf z),
    \qquad
    s=1,\ldots,n_s,
    \label{eq:tempered-bayesian-update}
\end{equation}
where $s$ denotes the stage index,
$n_s$ is the total number of stages,
$\pi^{(0)}(\mathbf z)=\pi^{\mathrm f}(\mathbf z)$,
and $\alpha_s>0$ is the tempering coefficient at stage $s$.
The tempering coefficients satisfy
$\sum_{s=1}^{n_s}\alpha_s=1$,
such that the final-stage distribution corresponds to the filtering posterior,
$\pi^{(n_s)}(\mathbf z)=\pi^{\mathrm a}(\mathbf z)$. The number of stages $n_s$ and the tempering coefficients $\{\alpha_s\}_{s=1}^{n_s}$ are treated as hyperparameters, with the specific values used in our experiments reported in Sec.~\ref{sec:lorenz96}.

At each stage, the generative filtering procedure and EnKF fallback mechanism described in Sec.~\ref{sec:EnGF} (Algorithm~\ref{alg:engf}) are applied using the tempered likelihood
$p(\mathbf d\mid\mathbf z)^{\alpha_s}$,
with the analysis ensemble from the preceding stage serving as the input ensemble for the next stage.
By incorporating the observation information progressively, the tempered EnGF mitigates particle-weight concentration and reduces the frequency of EnKF fallback due to insufficient ESS.

\subsection{Latent EnGF}
\label{sec:latent-EnGF}

Direct application of the EnGF to many physical systems is challenging because spatial discretization can lead to high-dimensional states, in which the GMM representation of the filtering prior cannot be reliably fitted from a moderate forecast ensemble (Sec.~\ref{sec:EnGF}).
However, many such systems admit low-dimensional latent representations that can be learned effectively using machine-learning techniques.
For example, turbulent-flow systems with state dimensions up to $O(10^6)$ have been represented using only 10--20 latent variables~\cite{solera2024beta,steinbrenner2026turbulence}.
This motivates performing the EnGF analysis in a low-dimensional latent space, where the GMM can instead be fitted reliably.

Here, we consider a decoder-based dimensionality-reduction approach similar to those developed in previous studies~\cite{bojanowski2017optimizing,park2019deepsdf,chandravamsi2026feature}.
Specifically, the $i$th forecast particle is reconstructed as
\begin{equation}
    \widehat{\mathbf z}_i^{\mathrm f}(\bm x)
    =
    \mathcal D
    \left(
        \bm x,
        \boldsymbol{\xi}_i^{\mathrm f};
        \boldsymbol{\theta}
    \right),
    \qquad
    i=1,\ldots,n_e,
    \label{eq:latent-decoder}
\end{equation}
where $\mathcal D$ is a neural-network decoder with parameters
$\boldsymbol{\theta}$ shared by all forecast particles,
$\bm x$ denotes the spatial coordinate,
$\boldsymbol{\xi}_i^{\mathrm f}\in\mathbb R^{n_\xi}$ is the latent representation of the $i$th forecast particle, with $n_\xi\ll n_z$,
and $\widehat{\mathbf z}_i^{\mathrm f}$ is the corresponding reconstructed state.

At each DA cycle, given the forecast ensemble
$\{\mathbf z_i^{\mathrm f}\}_{i=1}^{n_e}$,
the latent representations
$\{\boldsymbol{\xi}_i^{\mathrm f}\}_{i=1}^{n_e}$
and the shared decoder parameters $\boldsymbol{\theta}$ are jointly optimized by minimizing
\begin{equation}
    \mathcal L
    =
    \frac{1}{n_e}
    \sum_{i=1}^{n_e}
    \left\|
        \mathbf z_i^{\mathrm f}
        -
        \widehat{\mathbf z}_i^{\mathrm f}
    \right\|_2^2
    +
    \frac{\lambda_\xi}{n_e}
    \sum_{i=1}^{n_e}
    \left\|
        \boldsymbol{\xi}_i^{\mathrm f}
    \right\|_2^2,
    \label{eq:latent-reconstruction-loss}
\end{equation}
where $\lambda_\xi$ is the regularization parameter for the latent representations.
The optimized latent representations
$\{\boldsymbol{\xi}_i^{\mathrm f}\}_{i=1}^{n_e}$
constitute the latent forecast ensemble.

The EnGF analysis is then performed directly on the latent forecast ensemble, exactly as in Sec.~\ref{sec:EnGF}: the GMM is fitted to the latent forecast codes $\{\boldsymbol{\xi}_i^{\mathrm f}\}_{i=1}^{n_e}$, and additional latent particles are generated to form the augmented latent set $\{\widetilde{\boldsymbol{\xi}}_j\}_{j=1}^{n_p}$.
Evaluating the weights requires mapping each augmented latent particle to the observation space through the decoder $\mathcal D$ (with its trained parameters fixed) and the observation operator $\mathcal H$, so that Eq.~\eqref{eq:engf-unnormalized-weights} becomes
\begin{equation}
    \widetilde w_j
    \propto
    \exp\!\left[
        -\frac{1}{2}
        \left(\mathbf d-\mathcal H(\mathcal D(\widetilde{\boldsymbol{\xi}}_j))\right)^\top
        \mathbf R^{-1}
        \left(\mathbf d-\mathcal H(\mathcal D(\widetilde{\boldsymbol{\xi}}_j))\right)
    \right],
    \qquad j=1,\ldots,n_p,
    \label{eq:latent-engf-weights}
\end{equation}
where $\mathcal D(\widetilde{\boldsymbol{\xi}}_j)$ denotes the decoded state of the $j$th augmented latent particle.
Although the composed operator $\mathcal H\circ\mathcal D$ is in general nonlinear, the latent analysis requires no modification, since the EnGF already assimilates observations through an arbitrary, possibly nonlinear observation operator (Sec.~\ref{sec:EnGF}).
Systematic resampling of the augmented set yields the latent analysis ensemble $\{\boldsymbol{\xi}_i^{\mathrm a}\}_{i=1}^{n_e}$, which is finally mapped through the decoder to recover the analysis ensemble $\{\mathbf z_i^{\mathrm a}\}_{i=1}^{n_e}$ in the physical state space.

\section{Numerical experiments}
\label{sec:result}

We evaluate the proposed EnGF through four test cases, each designed to examine a distinct aspect of the method.
The doubling map serves as a proof of concept for tracking strongly non-Gaussian filtering distributions.
The Lorenz-63 system, a widely used benchmark in DA, provides a systematic evaluation across a range of DA settings.
The Lorenz-96 system demonstrates the \textit{tempered EnGF}, showing how likelihood tempering improves filtering performance as observations become increasingly informative.
Finally, Toro's shock tube problem demonstrates the \textit{latent EnGF}, illustrating how dimensionality reduction enables the EnGF to handle high-dimensional physical systems with an effective low-dimensional representation.
For the first two test cases, a PF with a sufficiently large ensemble is used to provide a high-fidelity reference approximation of the filtering distribution; such a reference becomes computationally impractical for the Lorenz-96 and shock tube problems.

We use different metrics to assess the filtering performance.
For the doubling map, where the filtering posterior can be strongly non-Gaussian and bimodal, we use the Wasserstein-2 ($W_2$) distance to quantify the discrepancy between the estimated distribution $\mu$ and reference distributions $\mu_\dagger$.
Since the state is one-dimensional, it takes the closed form
\begin{equation}
W_2(\mu,\mu_\dagger)
=
\left[
\int_0^1
\left|
F_\mu^{-1}(q)-F_{\mu_\dagger}^{-1}(q)
\right|^2
\,\mathrm{d}q
\right]^{1/2},
\label{eq:w2-distance}
\end{equation}
where $F_\mu^{-1}$ and $F_{\mu_\dagger}^{-1}$ denote the corresponding quantile functions.

For the remaining test cases, which are multidimensional so that the quantile-based $W_2$ above is no longer directly available, we assess state-estimation accuracy and ensemble uncertainty using the root-mean-square error (RMSE) and ensemble spread, respectively, the standard performance metrics in DA.
At each DA cycle, the RMSE is defined using the Euclidean norm as
\begin{equation}
    \mathrm{RMSE}
    =
    \frac{1}{\sqrt{n_z}}
    \left\|
        \overline{\mathbf z}^{\mathrm a}
        -
        \mathbf z_\dagger
    \right\|_2,
    \qquad
    \overline{\mathbf z}^{\mathrm a}
    =
    \frac{1}{n_e}
    \sum_{i=1}^{n_e}
    \mathbf z_i^{\mathrm a},
    \label{eq:rmse}
\end{equation}
and the ensemble spread as
\begin{equation}
    \mathrm{Spread}
    =
    \sqrt{
        \frac{1}{n_z}
        \operatorname{tr}\!\left(\mathbf P^{\mathrm a}\right)
    },
    \qquad
    \mathbf P^{\mathrm a}
    =
    \frac{1}{n_e-1}
    \sum_{i=1}^{n_e}
    \left(
        \mathbf z_i^{\mathrm a}
        -
        \overline{\mathbf z}^{\mathrm a}
    \right)
    \left(
        \mathbf z_i^{\mathrm a}
        -
        \overline{\mathbf z}^{\mathrm a}
    \right)^\top,
    \label{eq:spread}
\end{equation}
where $\mathbf z_\dagger \in \mathbb{R}^{n_z}$ denotes the true state and $\mathbf P^{\mathrm a}$ is the analysis ensemble covariance.
The RMSE-to-spread ratio is used to assess ensemble calibration.
Specifically, comparable spread and RMSE (i.e., $\mathrm{Spread}/\mathrm{RMSE}\approx 1$) indicate that the uncertainty represented by the ensemble is consistent with the actual state-estimation error, suggesting a well-calibrated ensemble.
In contrast, a spread appreciably smaller or larger than the RMSE indicates an overconfident or conservative ensemble, respectively.

\subsection{Doubling map}

\subsubsection{Case setup}

We first consider a noisy doubling map~\cite{halmos2017lectures,bach2026learning}, a one-dimensional chaotic dynamical system that provides a simple setting for studying non-Gaussian and bimodal filtering distributions.
At each DA cycle, the state $z$ is propagated according to
\begin{equation}
    \mathcal{F}(z)=2z+\epsilon \pmod{1}, \qquad
    \epsilon\sim\mathcal{N}(0,\sigma_z^2),
    \label{eq:doubling-map}
\end{equation}
where $\sigma_z^2=10^{-4}$ denotes the process-noise variance, and the modulo operation maps the state back to $[0,1)$.

We consider a nonlinear observation operator $\mathcal{H}:\mathbb{R}\rightarrow\mathbb{R}$ given by $\mathcal{H}(z)=\cos(2\pi z)$.
Synthetic observations at each DA cycle are generated from the true state $z_\dagger$ according to
\begin{equation}
    d
    =
    \mathcal{H}(z_\dagger)
    +
    \eta,
    \qquad
    \eta
    \sim
    \mathcal{N}(0,\sigma_d^2),
    \label{eq:doubling-map-observation}
\end{equation}
where $\sigma_d^2=0.04$ is the observation-error variance.
Since $\mathcal{H}(z)=\mathcal{H}(1-z)$, the observation operator can give rise to bimodal filtering posteriors.

The true state is initialized as $z_\dagger^{0}=0.23$, while the particles in the initial ensemble are independently sampled as
$z_i^{0}\sim\mathcal{N}(0.2,0.1^2)$, $i=1,\ldots,n_e$, and wrapped onto $[0,1)$.
The experiment is performed for 180 DA cycles using an ensemble size of $n_e=200$.
A bootstrap PF with $n_e=10^4$ particles is used to provide a high-fidelity reference approximation of the filtering posterior.

\subsubsection{Numerical results}

\begin{figure}[!htb]
\centering
\includegraphics[width=0.99\textwidth]{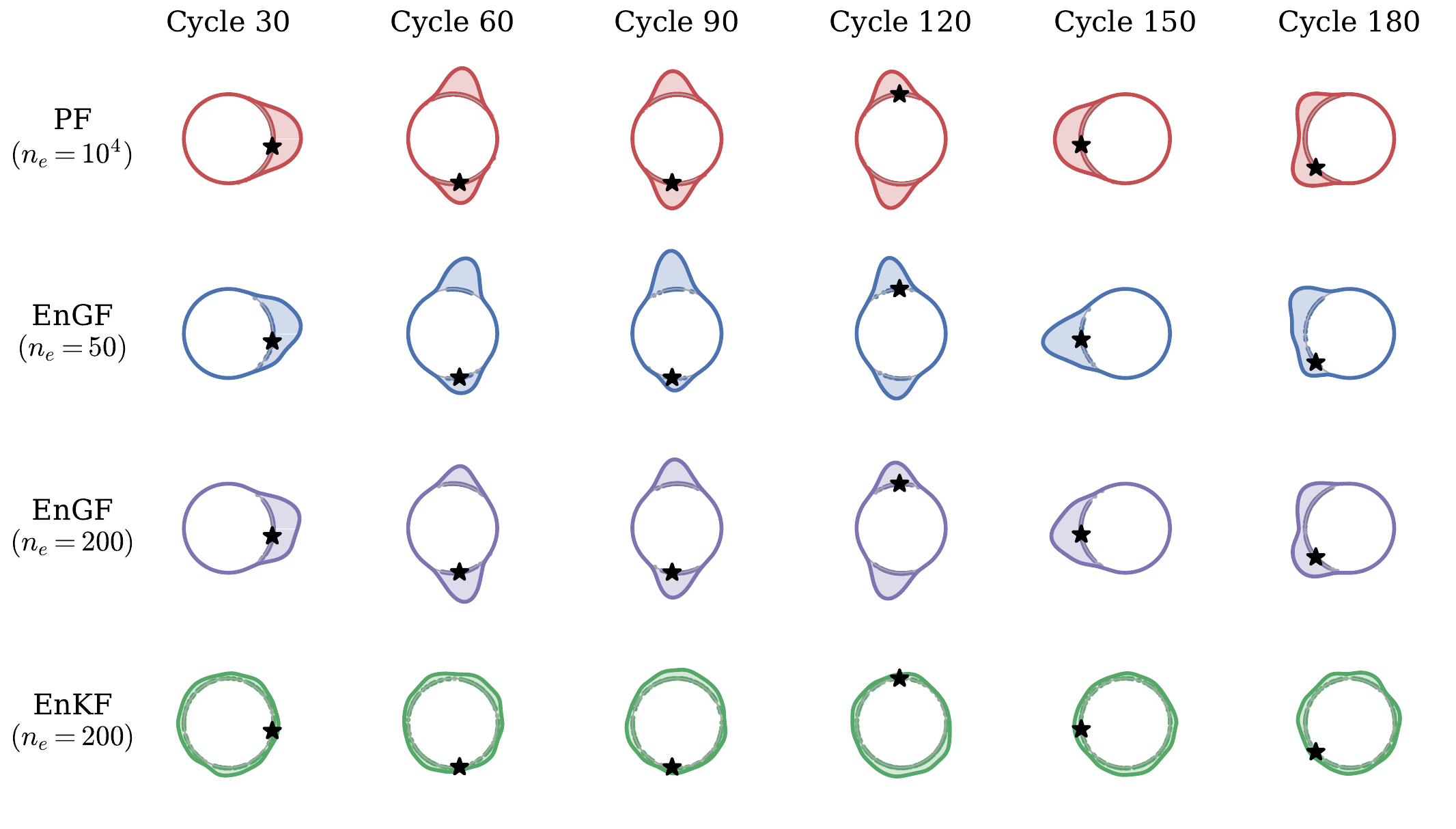}
\caption{Comparison of the filtering posteriors for the doubling map at selected DA cycles across different ensemble filters, showing that the EnGF captures the bimodal structure and closely matches the reference PF using a limited ensemble size, whereas the EnKF fails to resolve the two distinct modes.
The black stars (\blackstar{}) denote the true state.}
\label{fig:doubling-distribution}
\end{figure}

Fig.~\ref{fig:doubling-distribution} compares empirical approximations of the filtering posterior obtained by the EnGF and EnKF with the reference PF at selected DA cycles.
With the same ensemble size of $n_e=200$, the EnGF closely reproduces the reference filtering posterior, whereas the EnKF fails to resolve the two distinct modes.
Even with $n_e=50$, the EnGF retains the bimodal structure, although with reduced agreement with the reference filtering posterior.

\begin{figure}[!htb]
\centering
\includegraphics[width=0.55\textwidth]{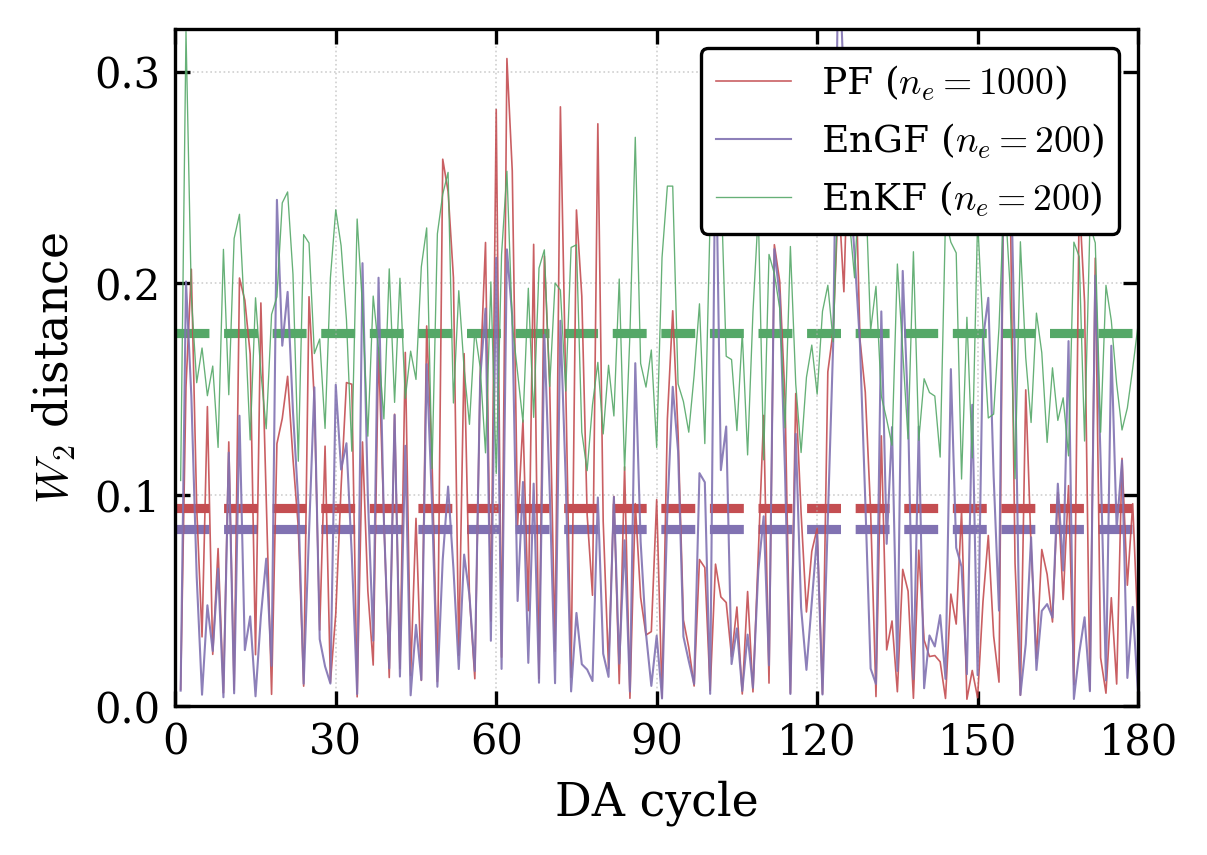}
    \caption{Estimated $W_2$ distance between the filtering posterior represented by each method and the PF reference ($n_e=10^4$) across DA cycles.
    Dashed lines denote the corresponding time-averaged $W_2$ distances over all 180 DA cycles.}
    \label{fig:doubling-w2}
\end{figure}

Fig.~\ref{fig:doubling-w2} further assesses the filtering performance by comparing the $W_2$ distance to the reference filtering posterior throughout the experiment.
With the same ensemble size of $n_e=200$, the EnGF consistently yields lower $W_2$ distances than the EnKF across the DA cycles and achieves a substantially lower time-averaged value (dashed lines).
The EnGF also outperforms the PF with $n_e=1000$, at only one-fifth of the forward-simulation cost.
For clarity, the EnGF with $n_e=50$ is not shown; its time-averaged $W_2$ distance lies between those of the EnGF with $n_e=200$ and the EnKF.

\subsection{Lorenz-63 system}

\subsubsection{Case setup}

We then consider the Lorenz-63 system~\cite{lorenz1963deterministic}, a three-dimensional chaotic dynamical system describing natural convection in a heated fluid and a canonical test case for sequential filtering under nonlinear dynamics.
The state $\mathbf{z}(t)=(z_1(t),z_2(t),z_3(t))$ evolves according to
\begin{equation}
    \frac{\mathrm{d}z_1}{\mathrm{d}t}
    = \sigma(z_2-z_1), \qquad
    \frac{\mathrm{d}z_2}{\mathrm{d}t}
    = z_1(\rho-z_3)-z_2, \qquad
    \frac{\mathrm{d}z_3}{\mathrm{d}t}
    = z_1z_2-\beta z_3,
    \label{eq:l63}
\end{equation}
where $\sigma=10$, $\rho=28$, and $\beta=8/3$.
Accordingly, the forecast operator $\mathcal{F}$ at each DA cycle is defined by integrating Eq.~\eqref{eq:l63} over the observation interval $\Delta t_{\mathrm{obs}}$.
Time integration is performed using the fourth-order Runge--Kutta method with a time step of $\Delta t=0.05$.

We first consider full observations of all three state variables, with
$\mathcal{H}:\mathbb{R}^{3}\rightarrow\mathbb{R}^{3}$ given by
$\mathcal{H}(\mathbf{z})=\mathbf{z}$.
Synthetic observations at each DA cycle are generated from the true state $\mathbf{z}_{\dagger}$ according to
\begin{equation}
    \mathbf{d}
    =
    \mathcal{H}(\mathbf{z}_{\dagger})
    +
    \boldsymbol{\eta},
    \qquad
    \boldsymbol{\eta}
    \sim
    \mathcal{N}(\mathbf{0},\sigma_d^2\mathbf{I}_3),
    \label{eq:l63-linear-observation}
\end{equation}
where $\sigma_d^2=4$ is the observation-error variance.

The true state is initialized as
$\mathbf{z}_\dagger^{0}=(1.508870,-1.531271,25.46091)$,
while the particles in the initial ensemble are independently sampled as
$\mathbf{z}_i^{0}\sim\mathcal{N}(\mathbf{0},4\mathbf{I}_3)$,
$i=1,\ldots,n_e$.
Each experiment consists of 1200 DA cycles: 200 spin-up cycles followed by 1000 cycles for evaluating the filtering performance.
We first consider a baseline setting with $\Delta t_{\mathrm{obs}}=0.1$ and $n_e=1000$, with $K_{\max}=8$ for the EnGF.
A bootstrap PF with $n_e=10^6$ particles is used to provide a high-fidelity reference for filtering performance.
To give a concrete sense of computational efficiency, we report timing for this baseline setting: the EnGF analysis requires an average wall-clock time of less than $0.1$~s per DA cycle using a CPU-based implementation, where parallel BIC evaluations ensure that the time remains nearly unchanged as $K_{\max}$ increases.

\subsubsection{Numerical results}

\begin{figure}[!htb]
\centering
\includegraphics[width=0.99\textwidth]{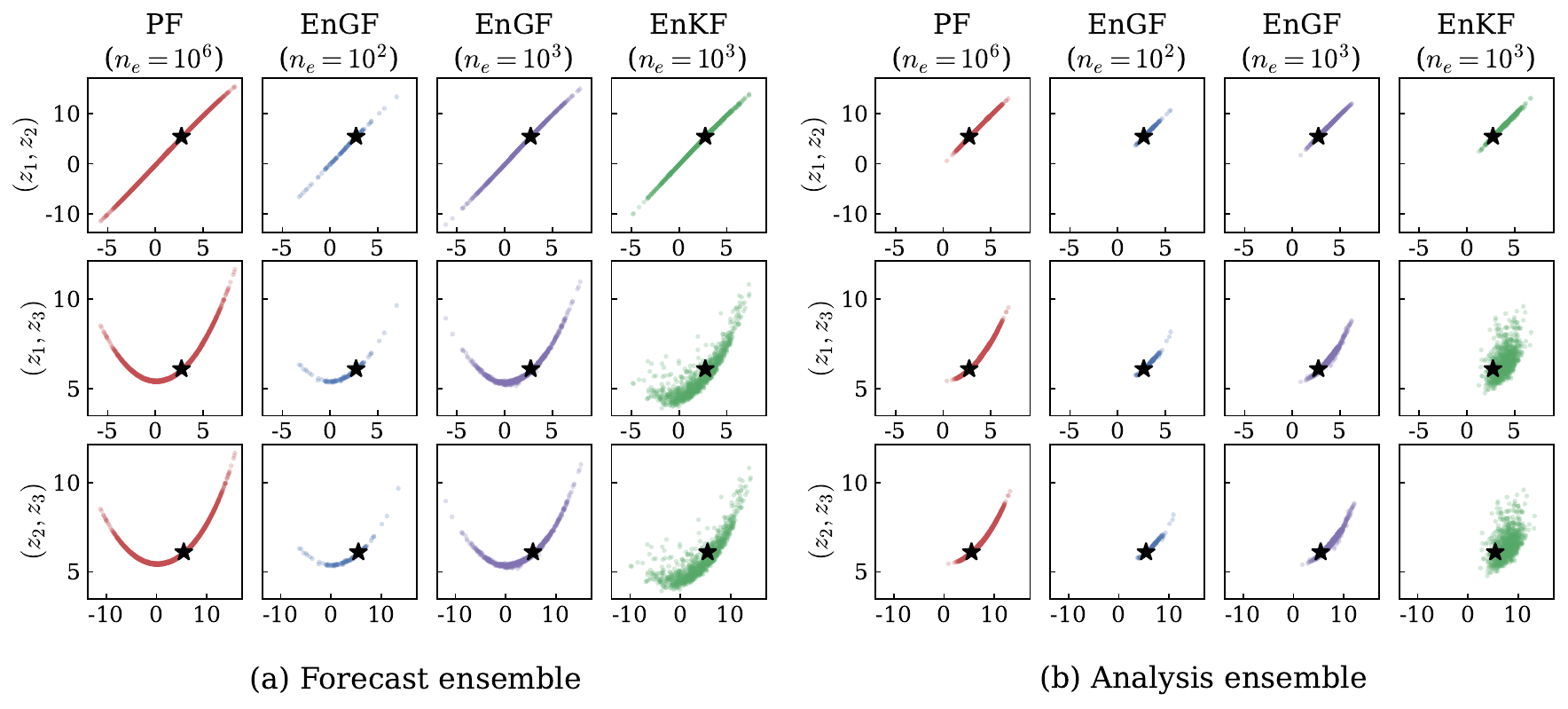}
    \caption{Forecast and analysis ensembles at a representative DA cycle for the Lorenz-63 system, showing that the EnGF effectively incorporates the prior information into the analysis and closely agrees with the reference PF, whereas the EnKF analysis differs markedly from the reference filtering posterior.
    The black stars (\blackstar{}) denote the true state.}
    \label{fig:l63-filtering}
\end{figure}

Fig.~\ref{fig:l63-filtering} shows the forecast and analysis ensembles obtained by the reference PF, EnGF, and EnKF at a representative DA cycle.
The reference PF forecast ensemble represents a strongly non-Gaussian filtering prior (left panel).
The EnGF with $n_e=1000$ exhibits a similar structure and effectively incorporates the prior information into the analysis, yielding a filtering posterior that closely agrees with the reference PF (right panel).
Even with $n_e=100$, the EnGF still captures the main features of the filtering prior and posterior represented by the reference PF.
The EnKF provides a reasonable but more diffuse representation of the filtering prior; however, its analysis does not fully utilize the non-Gaussian prior features due to the Gaussian approximation of the prior, resulting in a posterior that differs markedly from the reference PF.

\begin{figure}[!htb]
\centering
\includegraphics[width=0.99\textwidth]{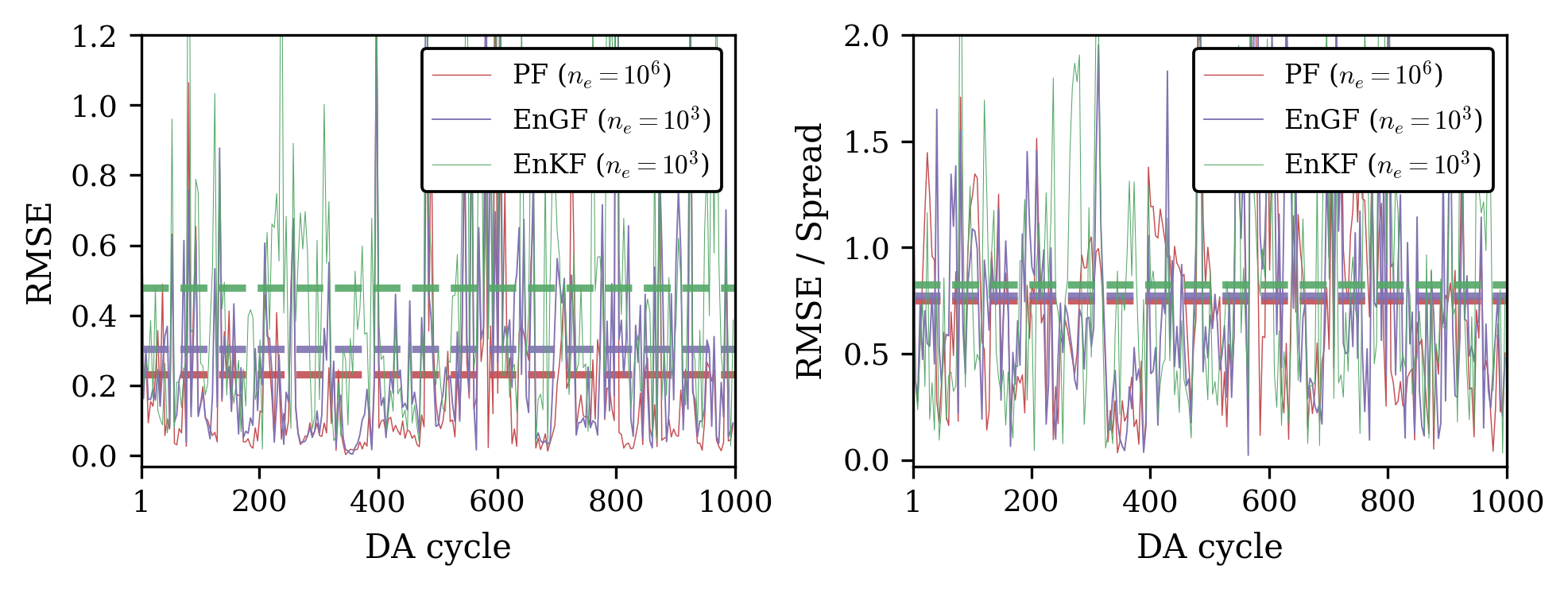}
    \caption{RMSE and RMSE-to-spread ratio over 1000 post-spin-up DA cycles for the Lorenz-63 system, showing substantially lower RMSE for the EnGF than the EnKF and performance approaching the reference PF (left), while all three methods maintain sufficient uncertainty relative to the state-estimation error (right).
    Dashed lines denote the corresponding 1000-cycle averages.}
    \label{fig:l63-rmse-vs-cycle}
\end{figure}

Fig.~\ref{fig:l63-rmse-vs-cycle} further evaluates the long-term filtering performance over the 1000 post-spin-up DA cycles using the RMSE and ensemble spread.
The EnGF achieves a markedly lower time-averaged RMSE than the EnKF and approaches the reference PF (left panel), as indicated by the horizontal dashed lines. 
The time-averaged RMSE-to-spread ratios remain around 0.8 for all three methods (right panel), indicating that the improved accuracy of the EnGF is achieved while maintaining sufficient uncertainty, without evidence of ensemble collapse.

\begin{figure}[!htb]
\centering
\includegraphics[width=0.99\textwidth]{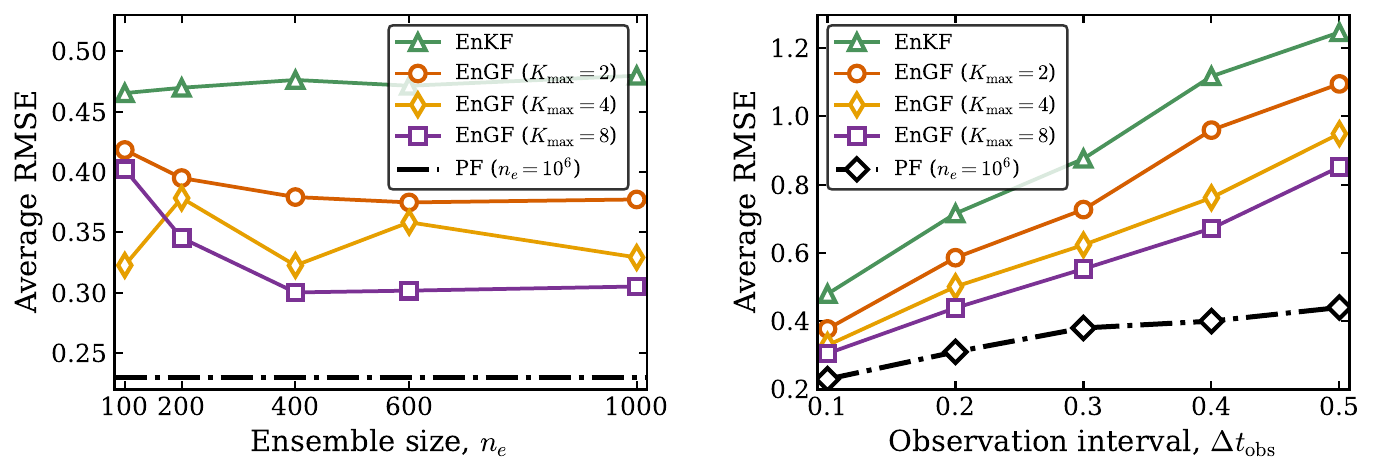}
    \caption{Time-averaged RMSE of the EnKF and EnGF under varying ensemble sizes (left panel, fixed $\Delta t_{\mathrm{obs}}=0.1$) and observation intervals (right panel, fixed $n_e=1000$), with the reference PF shown for comparison.
    For the EnGF, $K_{\max}$ denotes the prescribed upper bound on the number of GMM components.
    All RMSE values represent averages over 1000 post-spin-up DA cycles.}
    \label{fig:l63-rmse-sensitivity}
\end{figure}

We further investigate the effect of ensemble size on the long-term filtering performance at $\Delta t_{\mathrm{obs}}=0.1$, as shown in Fig.~\ref{fig:l63-rmse-sensitivity} (left panel).
The EnKF performance exhibits a clear plateau as the ensemble size increases, suggesting that, over the range considered here ($n_e=100$--$1000$), its error is primarily associated with the Gaussian approximation of the filtering prior rather than finite-ensemble sampling error.
In contrast, the EnGF generally achieves lower time-averaged RMSE as the ensemble size increases, consistent with an improved representation of the filtering prior as more particles become available.
Increasing $K_{\max}$ provides additional flexibility in representing the filtering prior, generally leading to lower time-averaged RMSE.
Notably, the best-performing EnGF configurations ($n_e \geq 400$ and $K_{\max}=8$) close nearly $80\%$ of the RMSE gap between the EnKF and the reference PF. 

We then examine the filtering performance as the observation interval $\Delta t_{\mathrm{obs}}$ increases, with the ensemble size fixed at $n_e=1000$, as shown in Fig.~\ref{fig:l63-rmse-sensitivity} (right panel).
Longer observation intervals allow nonlinear effects to accumulate during the forecast (i.e., increasing the effective nonlinearity of $\mathcal{F}$), generally producing more non-Gaussian filtering priors and a more challenging filtering problem.
Accordingly, the time-averaged RMSE increases for all methods as $\Delta t_{\mathrm{obs}}$ increases.
The EnGF consistently outperforms the EnKF, with its performance advantage increasing with $K_{\max}$, highlighting the benefit of greater flexibility in representing complex filtering priors.

\begin{figure}[!htb]
\centering
\includegraphics[width=0.55\textwidth]{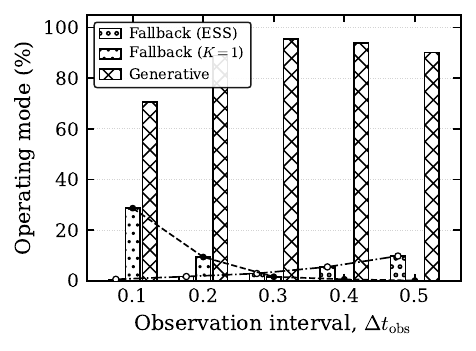}
    \caption{Occurrence frequencies of the EnGF operating modes across DA experiments with different observation intervals, showing that generative filtering remains dominant across all cases, while the $K=1$ fallback to EnKF becomes less frequent and the low-ESS fallback to EnKF becomes more frequent as $\Delta t_{\mathrm{obs}}$ increases.}
    \label{fig:l63-operating-mode}
\end{figure}

To understand the algorithmic behavior underlying this performance, Fig.~\ref{fig:l63-operating-mode} reports the occurrence frequencies of the three EnGF operating modes for the observation intervals considered above.
As introduced in Sec.~\ref{sec:EnGF}, the three modes consist of the standard generative filtering update and two EnKF fallback modes, triggered when either a single Gaussian component is selected ($K=1$) or the ESS is insufficient.
Fig.~\ref{fig:l63-operating-mode} shows that generative filtering remains dominant across all DA experiments, confirming it as the primary operating mode of the EnGF and providing a mechanistic explanation for its improved performance relative to the EnKF.
As $\Delta t_{\mathrm{obs}}$ increases, the $K=1$ fallback becomes less frequent, consistent with increasingly non-Gaussian filtering priors, whereas the low-ESS fallback becomes more frequent as weight degeneracy becomes more severe.
Overall, these results show that the EnGF appropriately switches between operating modes under different filtering conditions while maintaining generative filtering as the dominant mode.

Beyond the linear observation setting examined above, we also consider a nonlinear observation operator $\mathcal{H}(\mathbf{z})=\arctan(\mathbf{z})$, applied componentwise to all three state variables.
Synthetic observations are generated following Eq.~\eqref{eq:l63-linear-observation}, with observation-error variance $\sigma_d^2=0.1$, while the true initial state and initial ensemble remain unchanged.

\begin{figure}[!htb]
\centering
\includegraphics[width=0.99\textwidth]{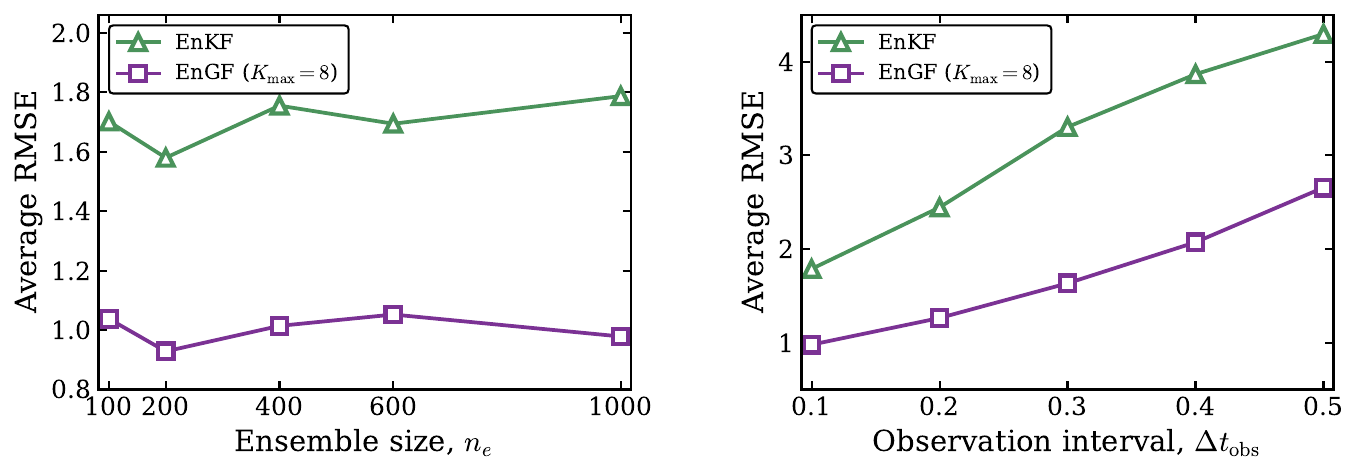}
    \caption{Time-averaged RMSE of the EnKF and EnGF under the nonlinear observation operator for varying ensemble sizes (left panel, fixed $\Delta t_{\mathrm{obs}}=0.1$) and observation intervals (right panel, fixed $n_e=1000$).
    For the EnGF, $K_{\max}=8$ is fixed throughout.
    All RMSE values represent averages over 1000 post-spin-up DA cycles.}
    \label{fig:l63-nonlinear}
\end{figure}

The comparison of the EnKF and EnGF ($K_{\max}=8$) under the nonlinear observation operator is illustrated in Fig.~\ref{fig:l63-nonlinear}.
The EnGF achieves lower time-averaged RMSE than the EnKF across all ensemble sizes considered at fixed $\Delta t_{\mathrm{obs}}=0.1$ (left panel) and across all observation intervals considered at fixed $n_e=1000$ (right panel).
The performance advantage of the EnGF becomes more pronounced as the observation interval increases.

\subsection{Lorenz-96 system}
\label{sec:lorenz96}

\subsubsection{Case setup}

We next consider the Lorenz-96 system, a higher-dimensional chaotic dynamical system widely used as a benchmark for filtering algorithms in numerical weather prediction~\cite{lorenz1996predictability}.
In this study, we consider a 20-dimensional system with the state
$\mathbf{z}(t)=(z_1(t),\ldots,z_{20}(t))$ evolving according to
\begin{equation}
    \frac{\mathrm{d}z_j}{\mathrm{d}t}
    =
    (z_{j+1}-z_{j-2})z_{j-1}
    - z_j + F,
    \qquad j=1,\ldots,20,
    \label{eq:l96}
\end{equation}
with periodic boundary conditions ($z_{-1}=z_{19}$, $z_0=z_{20}$, and $z_{21}=z_1$) and a constant forcing parameter $F=8$, chosen such that the system exhibits chaotic behavior.
Accordingly, the forecast operator $\mathcal{F}$ at each DA cycle is defined by integrating Eq.~\eqref{eq:l96} over the observation interval $\Delta t_{\mathrm{obs}}$.
Time integration is performed using the fourth-order Runge--Kutta method with a time step of $\Delta t=0.02$.

We consider partial observations of five equally spaced state variables,
$\{z_1,z_5,z_9,z_{13},z_{17}\}$, with
$\mathcal{H}:\mathbb{R}^{20}\rightarrow\mathbb{R}^{5}$ extracting these variables.
Synthetic observations at each DA cycle are generated from the true state $\mathbf{z}_{\dagger}$ according to
\begin{equation}
    \mathbf{d}
    =
    \mathcal{H}(\mathbf{z}_{\dagger})
    +
    \boldsymbol{\eta},
    \qquad
    \boldsymbol{\eta}
    \sim
    \mathcal{N}(\mathbf{0},\sigma_d^2\mathbf{I}_5),
    \label{eq:l96-observation}
\end{equation}
where $\sigma_d^2=0.5$ is the observation-error variance.

The true initial state is sampled as
$\mathbf{z}_\dagger^{0}\sim\mathcal{N}(\mathbf{0},\mathbf{I}_{20})$,
while the particles in the initial ensemble are independently sampled as
$\mathbf{z}_i^{0}\sim\mathcal{N}(\mathbf{0},\mathbf{I}_{20})$,
$i=1,\ldots,n_e$.
Each experiment consists of 1200 DA cycles: 200 spin-up cycles followed by 1000 cycles for evaluating the filtering performance.
The observation interval is set to $\Delta t_{\mathrm{obs}}=0.2$, corresponding to approximately one day under the conventional atmospheric time scaling of the Lorenz-96 model, for which $0.05$ time units correspond to approximately 6~hours~\cite{lorenz1996predictability}.
The ensemble size is set to $n_e=1000$.
Here, we consider two-stage likelihood tempering with different choices of tempering factors $[\alpha_1, \alpha_2]$.
The tempering factors $[0.3,0.7]$, which yield the best filtering performance among the choices considered, are used for the primary filtering results presented below.
The sensitivity to the tempering factors is also examined.
For the EnKF, neither covariance localization nor inflation is used, as both have negligible effects for the ensemble size $n_e=1000$.

\subsubsection{Numerical results}

\begin{figure}[!htb]
\centering
\includegraphics[width=0.99\textwidth]{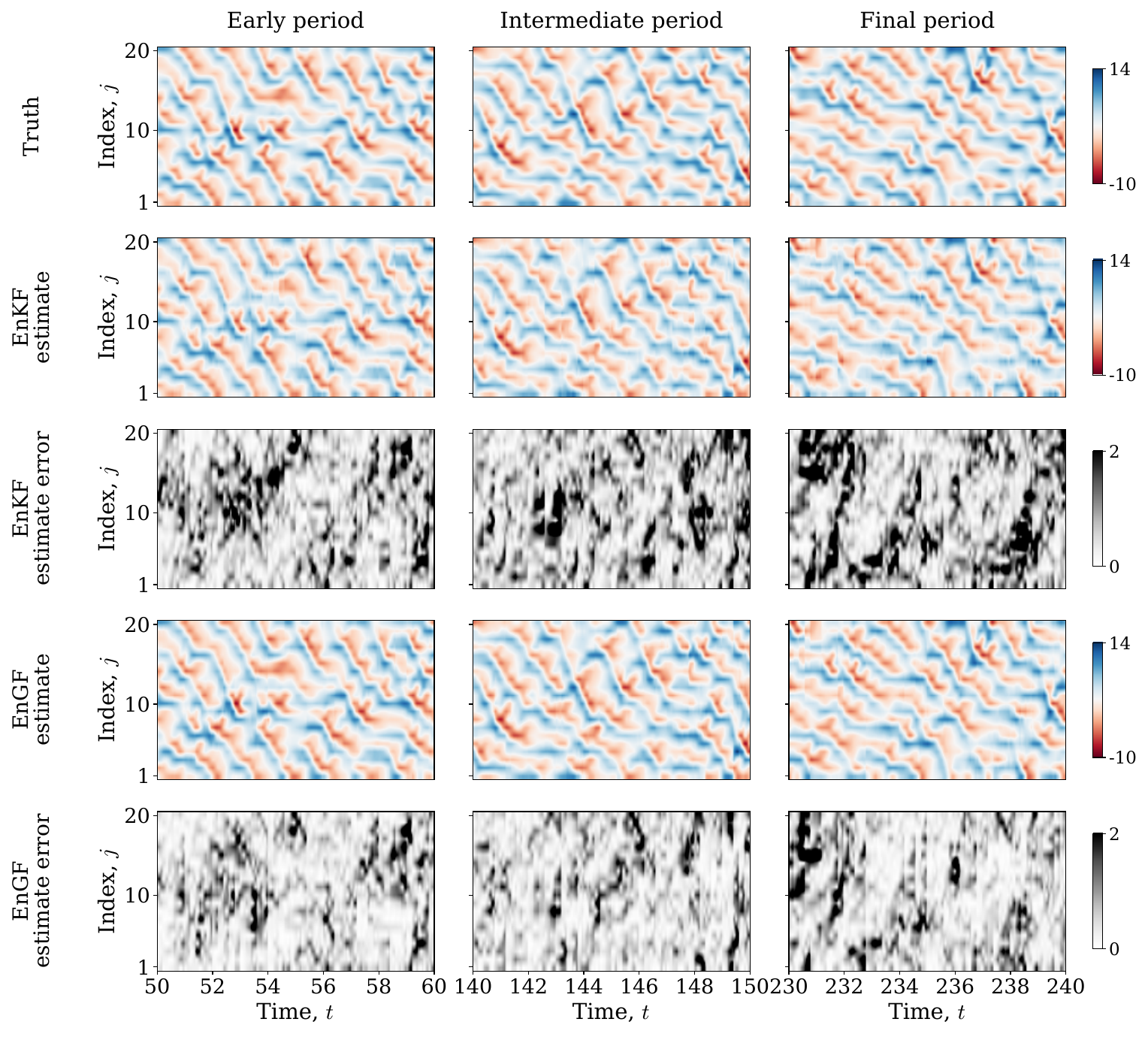}
    \caption{Comparison of the state estimates and corresponding errors obtained with the EnKF and EnGF for the Lorenz-96 system during three periods selected from the 1000 post-spin-up DA cycles ($t=40$--$240$): early, intermediate, and final, with each period spanning 50 cycles. 
    The state estimates correspond to the ensemble means, and the errors are reported as absolute errors relative to the truth.}
    \label{fig:l96-state-compare}
\end{figure}

Fig.~\ref{fig:l96-state-compare} compares the ensemble-mean state estimates obtained with the EnKF and EnGF during the early, intermediate, and final periods of the post-spin-up DA experiment, with each period showing the state evolution over 50 consecutive DA cycles.
Both methods reasonably track the spatiotemporal evolution of the true state throughout the three periods, despite the chaotic dynamics and partial observations.
However, the EnGF estimates remain consistently closer to the truth, exhibiting smaller absolute errors than the EnKF across all three periods.

\begin{figure}[!htb]
\centering
\includegraphics[width=0.99\textwidth]{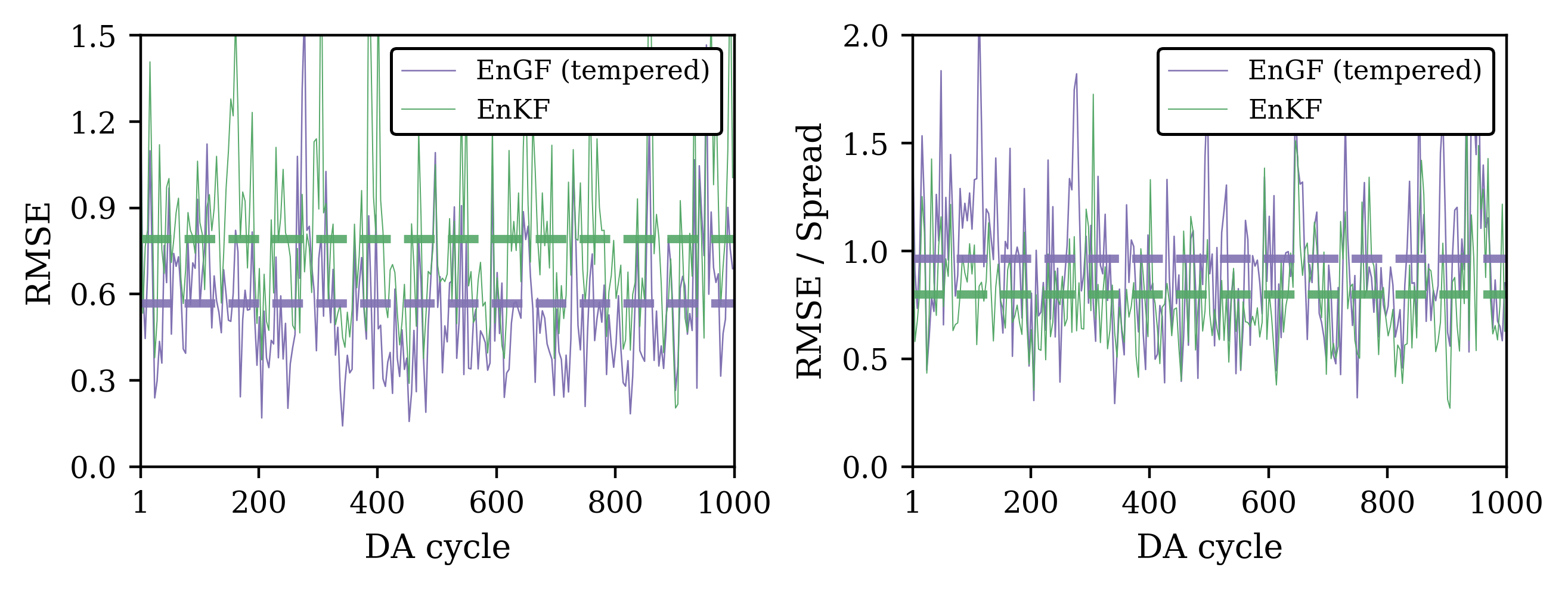}
    \caption{Comparison of the filtering performance of the EnKF and tempered EnGF for the Lorenz-96 system over 1000 post-spin-up DA cycles, showing lower RMSE for the EnGF (left panel) and closer agreement between RMSE and ensemble spread for the EnGF (right panel). 
    Dashed horizontal lines indicate the corresponding 1000-cycle averages.}
    \label{fig:l96-rmse-vs-cycle}
\end{figure}

The difference in filtering performance is further quantified by comparing the RMSE and ensemble spread of the EnKF and EnGF over the 1000 post-spin-up DA cycles, as shown in Fig.~\ref{fig:l96-rmse-vs-cycle}.
The tempered EnGF consistently achieves lower RMSE than the EnKF (left panel), resulting in a substantially lower time-averaged RMSE over the entire experiment.
Despite this improved accuracy, the RMSE and ensemble spread remain closely matched for the EnGF (right panel), indicating a well-calibrated estimate of the uncertainty over the entire experiment.

\begin{figure}[!htb]
\centering
\includegraphics[width=0.99\textwidth]{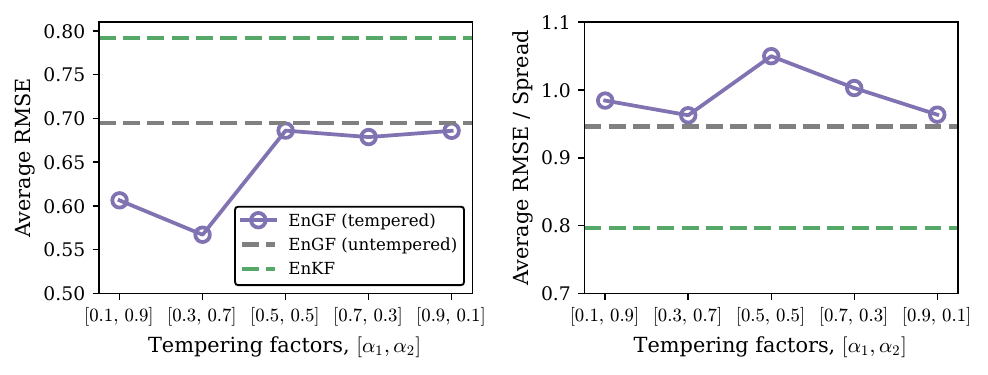}
    \caption{Sensitivity of the tempered EnGF to the choice of tempering factors for the Lorenz-96 system, showing consistently lower time-averaged RMSE than the untempered EnGF and EnKF for all tested tempering factors (left panel), while maintaining comparable RMSE and ensemble spread (right panel). 
    Dashed horizontal lines denote the corresponding 1000-cycle averages for the untempered EnGF and EnKF.}
    \label{fig:l96-temper}
\end{figure}

The sensitivity of the EnGF to the choice of tempering factors is further examined in Fig.~\ref{fig:l96-temper}.
Across all tested tempering choices, the tempered EnGF consistently achieves a lower time-averaged RMSE than both the untempered EnGF and EnKF (left panel).
Among the tempered configurations, smaller first-stage tempering factors generally lead to lower RMSE.
Meanwhile, the RMSE and ensemble spread remain closely matched in all settings (right panel), indicating reasonable estimates of the state uncertainty across the different tempering choices.

The improved filtering performance observed for smaller first-stage tempering factors can be understood from the ESS-based fallback mechanism in the EnGF.
When the first-stage tempering factor $\alpha_1$ is large, the importance weights tend to become more concentrated, resulting in a lower ESS.
If the ESS falls below the prescribed threshold, the EnGF reverts to the EnKF update in the first stage, in which the filtering prior is approximated as Gaussian.
Consequently, non-Gaussian features of the filtering prior represented by the GMM, such as multimodality and skewness, cannot be fully exploited in the analysis.
In contrast, a smaller first-stage tempering factor produces a more moderate likelihood weighting, which helps maintain a sufficiently large ESS and therefore allows the EnGF to retain and exploit the distributional structure of the filtering prior.
Moreover, because the intermediate ensemble produced by the first-stage analysis serves as the prior for the second stage, preserving this distributional information in the first stage can also benefit the second-stage EnGF update.

\subsection{Toro's shock tube problem}

\subsubsection{Case setup}

We finally consider Toro's shock tube problem~\cite{toro2013riemann}, a benchmark compressible-flow problem that provides a more physically representative test for high-dimensional, PDE-based systems.
Beyond its high dimensionality, it is a demanding filtering test because uncertainty in the shock location places forecast particles on opposite sides of the discontinuity, making the filtering prior near the shock bimodal and strongly non-Gaussian~\cite{zhou2026neural,chandravamsi2026feature}.
A Gaussian, linear analysis then superposes these displaced discontinuities, smearing sharp features and generating spurious oscillations and nonphysical states, so the shock tube problem is a stringent non-Gaussian benchmark for the latent EnGF.

The flow dynamics are governed by the compressible Euler equations,
\begin{subequations}
    \label{eq:euler}
    \begin{align}
        \frac{\partial \rho}{\partial t}
        + \nabla \cdot (\rho \bm{V}) &= 0, \\
        \frac{\partial (\rho \bm{V})}{\partial t}
        + \nabla \cdot (\rho \bm{V} \otimes \bm{V})
        + \nabla p &= 0, \\
        \frac{\partial E}{\partial t}
        + \nabla \cdot \left[(E+p)\bm{V}\right] &= 0.
    \end{align}
\end{subequations}
Here, $\rho$, $\bm{V}=[u,v,w]^\top$, and $p$ denote the density, velocity, and pressure, respectively, and $\otimes$ denotes the outer product.
The total energy per unit volume $E$ is given by
\begin{equation}
E = \rho\left(\frac{\bm{V}\cdot\bm{V}}{2} + e\right)
= \rho\left(\frac{u^2+v^2+w^2}{2} + e\right),
\label{eq:toro-total-energy}
\end{equation}
where $e$ is the specific internal energy.
Assuming a perfect gas, the system is closed by the equation of state
\begin{equation}
p = (\gamma-1)\rho e,
\label{eq:toro-eos}
\end{equation}
where $\gamma=1.4$ is the ratio of specific heats.
In the one-dimensional setting considered here, $\bm V$ reduces to the scalar velocity $u$.

The computational domain $x\in[0,1]$ is discretized using a uniform grid with $n_x=400$ cells.
Accordingly, the state vector, formed by the flow variables $\rho$, $u$, and $p$ evaluated on the grid, is given by
\begin{equation}
\mathbf{z}(t)
=
\big(
\rho_1(t),\ldots,\rho_{n_x}(t),
u_1(t),\ldots,u_{n_x}(t),
p_1(t),\ldots,p_{n_x}(t)
\big),
\label{eq:toro-state}
\end{equation}
with state dimension $n_z=3n_x=1200$.
Non-reflecting boundary conditions are imposed at both ends of the domain.
At each DA cycle, the forecast operator $\mathcal{F}$ advances the state over the observation interval $\Delta t_\mathrm{obs}$ by numerically integrating the one-dimensional form of Eq.~\eqref{eq:euler}.
The numerical integration is performed using the open-source solver \texttt{M2C} (Multiphysics Modeling and Computation)~\cite{zhao2026m2c,wang_m2c}, which employs a MUSCL scheme~\cite{van1979towards}.
Numerical fluxes are computed using a local Lax--Friedrichs approximate Riemann solver, and time integration is performed using an explicit second-order Runge--Kutta method with a time step of $\Delta t=2\times10^{-5}$.

We consider sparse pressure observations at five spatial locations, $x=0.1$, $0.3$, $0.5$, $0.7$, and $0.9$, where $\mathcal{H}:\mathbb{R}^{1200}\rightarrow\mathbb{R}^{5}$ extracts the pressure at these locations.
Synthetic observations at each DA cycle are generated from the true state $\mathbf{z}_{\dagger}$ according to
\begin{equation}
\mathbf{d}
=
\mathcal{H}(\mathbf{z}_{\dagger})
+
\boldsymbol{\eta},
\qquad
\boldsymbol{\eta}
\sim
\mathcal{N}(\mathbf{0},\mathbf{R}),
\label{eq:toro-observation}
\end{equation}
where the observation-error covariance matrix $\mathbf{R}$ is diagonal, with entries
\begin{equation}
R_{jj}
=
\left(
0.05[\mathcal{H}(\mathbf{z}_{\dagger})]_j
+ 5.0
\right)^2,
\qquad j=1,\ldots,5.
\label{eq:toro-observation-covariance}
\end{equation}

The initial condition consists of two constant states separated by a diaphragm located at $x=x_d$,
\begin{equation}
    (\rho,u,p)
    =
    \begin{cases}
        (\rho_L,u_L,p_L), & x<x_d,\\
        (\rho_R,u_R,p_R), & x\geq x_d.
    \end{cases}
    \label{eq:toro-initial-condition}
\end{equation}
Table~\ref{tab:toro-initial-condition} summarizes the true values of the initial-condition parameters. The initial ensemble is generated by sampling these parameters from Gaussian distributions with the means and standard deviations also listed in the table.
The resulting initial ensemble, together with the true initial state, is shown in the top row of Fig.~\ref{fig:toro-profile-compare}.

\begin{table}[!htb]
\centering
\caption{Parameters used to define the true initial condition and generate
the initial ensemble for Toro's shock tube problem.
$(\rho_L,u_L,p_L)$ and $(\rho_R,u_R,p_R)$ denote the states to the left
and right of the diaphragm, respectively, and $x_d$ denotes the diaphragm
location. Each parameter is sampled from a Gaussian with the mean and standard deviation listed in the table to generate the initial ensemble, with the mean coinciding with the true initial condition.}
\label{tab:toro-initial-condition}
\begin{tabular}{lccccccc}
\toprule
& $\rho_L$ & $u_L$ & $p_L$
& $\rho_R$ & $u_R$ & $p_R$ & $x_d$ \\
\midrule
Truth / mean
& 5.99924 & 19.5975 & 460.894
& 5.99242 & $-6.19633$ & 46.0950 & 0.4 \\
Standard deviation
& 0.5 & 0.0 & 20.0
& 0.0 & 0.0 & 2.0 & 0.15 \\
\bottomrule
\end{tabular}
\end{table}

Due to the rapid evolution and short characteristic time scale of the shock-tube dynamics, we use a small observation interval of $\Delta t_{\mathrm{obs}}=7\times10^{-3}$ and perform DA over five cycles using an ensemble of $n_e=100$ particles.
Despite the relatively small number of DA cycles, this time window captures substantial evolution of the flow.
At each DA cycle, the decoder is trained to represent each forecast particle $\mathbf{z}^\mathrm{f}_i\in\mathbb{R}^{1200}$ through a low-dimensional latent representation $\boldsymbol{\xi}^\mathrm{f}_i\in\mathbb{R}^{6}$.
The EnGF analysis is then performed in this latent space following the procedure described in Sec.~\ref{sec:latent-EnGF}, and each resulting latent analysis particle $\boldsymbol{\xi}^\mathrm{a}_i$ is mapped through the trained decoder to obtain the corresponding analysis particle $\mathbf{z}^\mathrm{a}_i$ in the physical space.
For comparison, the EnKF is also applied in the same latent space; the standard EnKF update in the physical space is not considered because it can produce nonphysical states (i.e., negative density and pressure) and spurious oscillations, causing the subsequent forecast simulation to fail~\cite{zhou2026neural}.

\subsubsection{Numerical results}

\begin{figure}[!htb]
\centering
\includegraphics[width=0.99\textwidth]{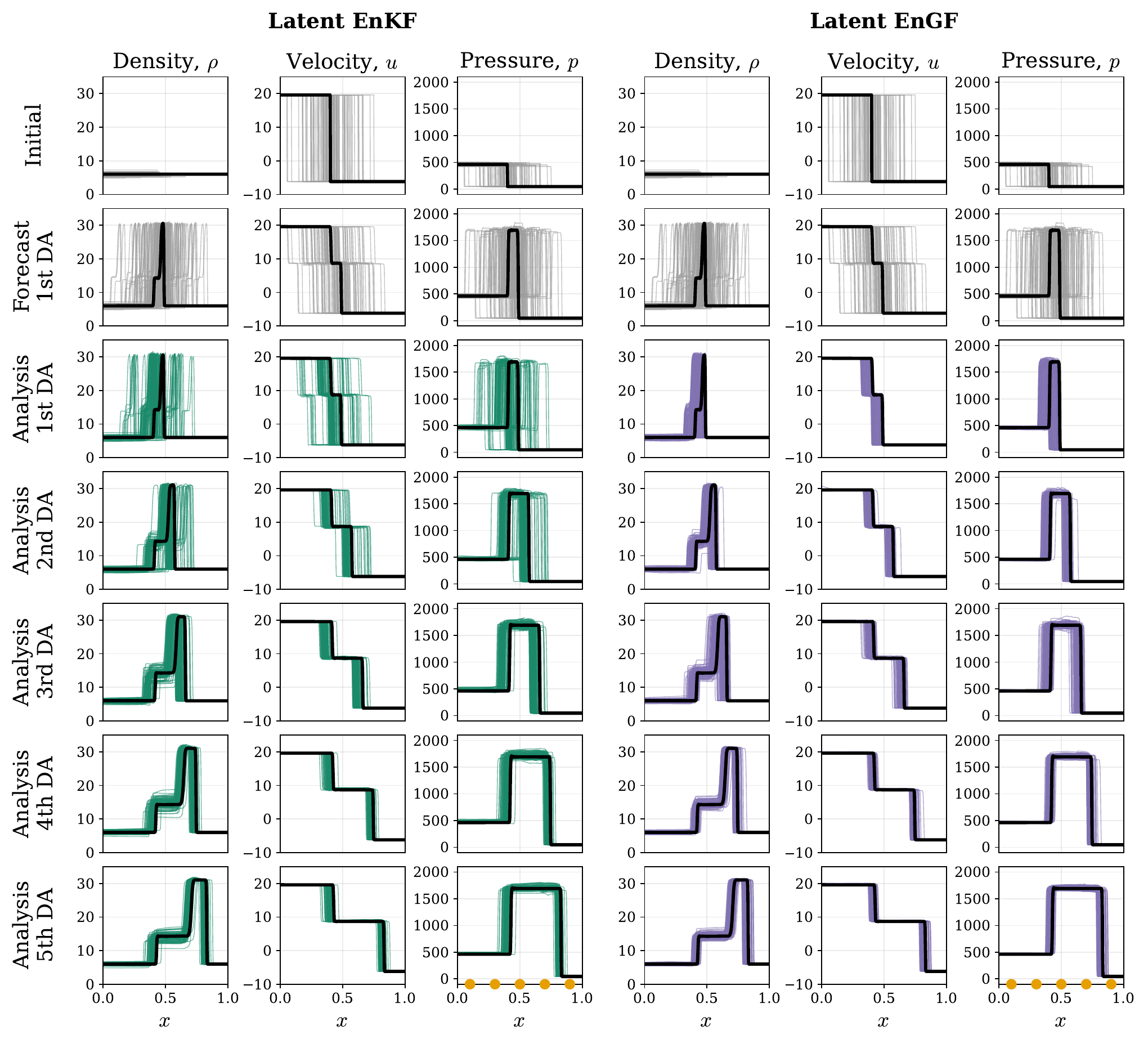}
    \caption{Comparison of latent EnKF (left) and latent EnGF (right) for Toro's shock tube problem, showing more rapid concentration of the EnGF analysis ensembles around the true solution over successive DA updates.
    The rows show the initial ensemble, the forecast ensemble before the first DA update, and the analysis ensembles after the first through fifth DA updates.
    Black curves denote the true solution, colored curves represent particles, and orange markers indicate the pressure sensor locations ($x=0.1$, $0.3$, $0.5$, $0.7$, and $0.9$).}
    \label{fig:toro-profile-compare}
\end{figure}

Fig.~\ref{fig:toro-profile-compare} compares the latent EnKF and the latent EnGF ensembles over five DA cycles, showing that the latent EnGF produces analysis ensembles that converge more rapidly towards the true solution.
Note that the latent EnKF propagates the same set of particles (green) across DA cycles, whereas the particle set in the latent EnGF (purple) can change after each analysis step through the generation and resampling of new particles.
Starting from the same initial ensemble (top row), both methods exhibit substantial uncertainty in the first forecast (second row), particularly near sharp flow features.
After the first DA update (third row), the latent EnGF ensemble becomes markedly more concentrated around the true solution, while the latent EnKF retains considerably greater spread.
This difference persists over subsequent DA cycles, with the latent EnGF more closely tracking the true solution across all three flow variables.
Notably, although observations are available only for pressure at five spatial locations, the latent EnGF also substantially reduces uncertainty in density and velocity, demonstrating that assimilation in the latent space effectively transfers information from sparse pressure observations to the unobserved flow variables.

\begin{figure}[!htb]
\centering
\includegraphics[width=0.99\textwidth]{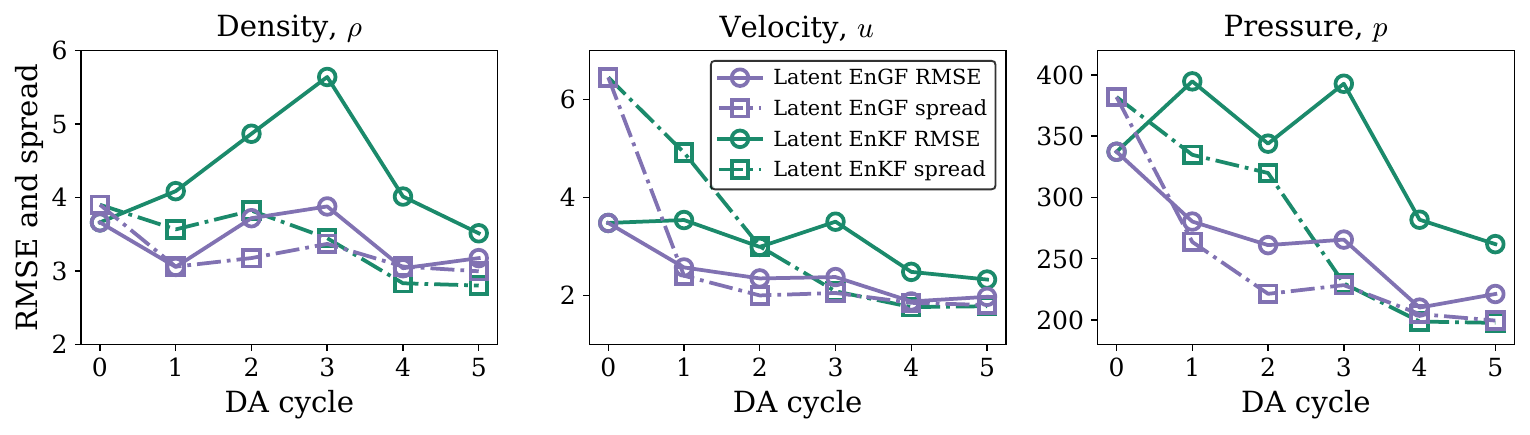}
    \caption{Comparison of RMSE and ensemble spread between the latent EnKF and latent EnGF across DA cycles for the three flow variables $(\rho, u, p)$ in Toro's shock tube problem, showing consistently lower RMSE and closer agreement between RMSE and ensemble spread for the latent EnGF.
    Cycle 0 corresponds to the forecast ensemble immediately before the first DA update, while cycles 1--5 correspond to the analysis ensembles after the first through fifth DA updates.}
    \label{fig:toro-rmse-spread}
\end{figure}

The performance difference is further quantified by the RMSE and ensemble spread shown in Fig.~\ref{fig:toro-rmse-spread}.
Cycle 0 corresponds to the forecast ensemble immediately before the first DA update, while cycles 1--5 correspond to the analysis ensembles after the first through fifth updates.
Across all five DA cycles and all three flow variables, the latent EnGF consistently achieves lower RMSE than the latent EnKF, demonstrating improved state-estimation accuracy.
The latent EnGF also exhibits closer agreement between RMSE and ensemble spread, indicating better consistency between the actual estimation error and the uncertainty represented by the ensemble.
Overall, these results demonstrate that the latent EnGF can effectively perform DA in high-dimensional flow systems that admit a suitable low-dimensional representation.

\section{Conclusion}
\label{sec:conclusion}

In this work, we introduced the EnGF, a simple yet effective generative filtering approach for exploiting non-Gaussian information contained in moderately large forecast ensembles.
The EnGF adaptively models the filtering prior using a GMM as a lightweight generative model and generates a much larger particle population for Bayesian analysis without requiring additional expensive model forecasts.

Across all test cases, the EnGF consistently outperforms the EnKF at the same forward-simulation cost, achieving lower state-estimation errors while maintaining adequate ensemble spread.
When reference PF solutions are available, the EnGF achieves filtering accuracy close to that of the reference PF while requiring orders of magnitude fewer forward simulations.
The Lorenz-63 results further show that the EnGF continues to benefit from increasing ensemble size after the EnKF performance largely plateaus.
This behavior indicates that, as sampling error decreases with increasing ensemble size, a more flexible representation of the filtering prior can mitigate the representation error associated with the Gaussian approximation.
The tempered and latent extensions further demonstrate that the same basic filtering strategy can accommodate informative observations and high-dimensional physical systems.
Taken together, these results position the EnGF as a practical intermediate between the EnKF and PFs for non-Gaussian filtering with moderately large ensembles.

Important challenges remain.
Prior modeling from finite ensembles becomes increasingly difficult as the intrinsic dimension grows, while particle degeneracy remains a fundamental limitation under highly informative observations.
Although the tempered and latent extensions mitigate these difficulties, they do not eliminate them.
Future work should therefore explore more scalable strategies for prior modeling and particle analysis, including adaptive tempering and local mixture modeling for cycling priors.

\section*{Acknowledgments}
The authors thank Matthias Morzfeld for helpful discussions and Bohan Chen for suggesting the doubling map case.
The authors also thank Xuning Zhao for assistance with the \texttt{M2C} solver.
This work used computational resources provided by the Advanced Cyberinfrastructure Coordination Ecosystem: Services \& Support (ACCESS) program under allocation PHY250246 and was carried out on Anvil at Purdue University.

\section*{Code availability}

The source code used to produce the numerical results in this study will be made publicly available upon publication in the GitHub repository~\cite{zhou2026engfcode}.

\end{document}